\documentclass{article}

\PassOptionsToPackage{numbers, compress}{natbib}

\usepackage{xspace}
\usepackage{soul}
\usepackage{url}
\usepackage{graphicx}
\usepackage{amsmath}
\usepackage{amsfonts}
\usepackage{multirow}
\usepackage{amsthm}
\usepackage{booktabs}
\usepackage{epstopdf}
\usepackage{xcolor}
\usepackage{bbm}
\usepackage{dsfont}
\usepackage{setspace}
\usepackage{braket}
\usepackage{tablefootnote}
\usepackage[normalem]{ulem}
\usepackage{hyperref}
\usepackage{cleveref}
\usepackage{url}
\usepackage{makecell}
\usepackage{wrapfig}
\usepackage{sansmath}
\usepackage{float}
\usepackage{pifont}
\useunder{\uline}{\ul}{}
\usepackage{tcolorbox}
\usepackage{fancyvrb}
\usepackage{listings}
\useunder{\uline}{\ul}{}
\usepackage[textsize=tiny]{todonotes}
\usepackage[toc,page,header]{appendix}
\usepackage{minitoc}

\newcommand{\ourmethod}{\texttt{GFM-RAG}\xspace}
\graphicspath{{images/}}

\usepackage{tikz}
\usetikzlibrary{tikzmark}
\tikzset{mycircled/.style={circle,draw,inner sep=0.05em,line width=0.1em, scale=0.8}}
\usepackage{xcolor}
\definecolor{mypurple}{HTML}{6600CC}

\usepackage{amsmath,amsfonts,bm}

\def\eqref#1{equation~\ref{#1}}

\def\1{\bm{1}}

\def\vzero{{\bm{0}}}

\def\vq{{\bm{q}}}

\DeclareMathAlphabet{\mathsfit}{\encodingdefault}{\sfdefault}{m}{sl}
\SetMathAlphabet{\mathsfit}{bold}{\encodingdefault}{\sfdefault}{bx}{n}

\def\gA{{\mathcal{A}}}

\def\gD{{\mathcal{D}}}
\def\gE{{\mathcal{E}}}

\def\gG{{\mathcal{G}}}

\def\gL{{\mathcal{L}}}

\def\gN{{\mathcal{N}}}

\def\gR{{\mathcal{R}}}

\def\gT{{\mathcal{T}}}

\def\sR{{\mathbb{R}}}

\usepackage[final]{neurips_2025}

\usepackage[utf8]{inputenc} 
\usepackage[T1]{fontenc}    
\usepackage{url}            
\usepackage{booktabs}       
\usepackage{amsfonts}       
\usepackage{nicefrac}       
\usepackage{microtype}      
\usepackage{xcolor}         

\newcommand{\github}{\raisebox{-1.5pt}{\includegraphics[height=1em]{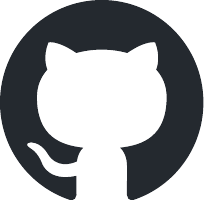}}}

\title{GFM-RAG: Graph Foundation Model for Retrieval Augmented Generation}

\author{Linhao Luo\textsuperscript{1}\thanks{Equal Contribution.},~Zicheng Zhao\textsuperscript{2}\footnotemark[1],~Gholamreza Haffari\textsuperscript{1},~Chen Gong\textsuperscript{3},~Dinh Phung\textsuperscript{1},~Shirui Pan\textsuperscript{4}\thanks{Corresponding author.}\\
\textsuperscript{1}Monash University,~\textsuperscript{2}Nanjing University of Science and Technology,\\\textsuperscript{3}Shanghai Jiao Tong University,~\textsuperscript{4}Griffith University,\\
\texttt{\{Linhao.Luo,gholamreza.haffari,dinh.phung\}@monash.edu,}\\\texttt{zicheng.zhao@njust.edu.cn,~chen.gong@sjtu.edu.cn,~s.pan@griffith.edu.au}\\
\hspace*{-0.8cm}\github{} \textbf{Project page:}~~\texttt{\url{https://rmanluo.github.io/gfm-rag}} \\
}

\begin{document}
\doparttoc 
\faketableofcontents 

\maketitle
\begin{abstract}
    Retrieval-augmented generation (RAG) has proven effective in integrating knowledge into large language models (LLMs). However, conventional RAGs struggle to capture complex relationships between pieces of knowledge, limiting their performance in intricate reasoning that requires integrating knowledge from multiple sources. Recently, graph-enhanced retrieval augmented generation (GraphRAG) builds graph structure to explicitly model these relationships, enabling more effective and efficient retrievers. Nevertheless, its performance is still hindered by the noise and incompleteness within the graph structure. To address this, we introduce \ourmethod, a novel graph foundation model (GFM) for retrieval augmented generation. \ourmethod is powered by an innovative graph neural network that reasons over graph structure to capture complex query-knowledge relationships. The GFM with 8M parameters undergoes a two-stage training process on large-scale datasets, comprising 60 knowledge graphs with over 14M triples and 700k documents. This results in impressive performance and generalizability for \ourmethod, making it the first graph foundation model applicable to unseen datasets for retrieval without any domain-specific fine-tuning required. Extensive experiments on three multi-hop QA datasets and seven domain-specific RAG datasets demonstrate that \ourmethod achieves state-of-the-art performance while maintaining efficiency and alignment with neural scaling laws, highlighting its potential for further improvement.
\end{abstract}

\section{Introduction}\label{sec:introduction}

Recent advancements in large language models (LLMs) \citep{gpt4o,llama3,qwen2} have greatly propelled the evolution of natural language processing, positioning them as foundational models for artificial general intelligence (AGI). Despite the remarkable reasoning ability \citep{openai_o1}, LLMs are still limited in accessing real-time information and lack of domain-specific knowledge, which is outside the pre-training corpus. To address these limitations, retrieval-augmented generation (RAG) \citep{gao2023retrieval} has become a popular paradigm in adding new knowledge to the static LLMs by retrieving relevant documents into the context of LLM generation. 

Existing RAG methods typically retrieve documents independently, making it difficult to capture complex relationships between pieces of knowledge \citep{karpukhin2020dense,bge_m3,moreira2024nv}. This limitation hampers the performance of LLMs in integrating knowledge across document boundaries, particularly in multi-hop reasoning tasks \citep{yang2018hotpotqa,trivedi2022musique} and real-world applications like legal judgment \citep{kang2024bridging} and medical diagnoses \citep{jin-etal-2019-pubmedqa}, which require reasoning over multiple sources. Although recent methods have expanded the retrieval process into multiple steps and incorporate LLM reasoning, they still encounter high computational costs due to iterative retrieval and reasoning with LLMs \citep{trivedi2023interleaving,sunthink,joshi2024reaper}.

Recently, graph-enhanced retrieval augmented generation (GraphRAG) \citep{peng2024graph,han2024retrieval} has emerged as a novel solution that builds a graph structure to explicitly model the intricate relationships between knowledge. This enables the development of a graph-enhanced retriever to identify relevant information using graphs. The structural nature of graphs allows GraphRAG to capture global context and dependencies among documents, significantly improving reasoning across multiple sources \citep{edge2024local}. 
Methods like HippoRAG \citep{gutiérrez2024hipporag} enhance retrieval by employing a personalized PageRank algorithm to locate relevant knowledge with graphs. However, these algorithms rely solely on the graph structure, which is often noisy or incomplete, limiting their overall performance.
Alternative methods \citep{mavromatis2024gnn,he2024g} incorporate graph neural networks (GNNs) into the retrieval process. These methods have shown impressive performance due to GNNs' powerful multi-hop reasoning capabilities on graphs \citep{yasunaga2021qa}. Nevertheless, they still face limitations in generalizability since they require training from scratch on new datasets.

Nowadays, the search for a foundation GNN model that can transfer and generalize across different datasets has been an active research topic. Ideally, a foundation GNN or graph foundation model (GFM) can benefit from large-scale training and generalize across diverse graphs \citep{maoposition,liu2025graph}. Efforts have been made to identify transferable graph tokens (e.g., motifs, sub-trees, and relation graphs) \citep{galkintowards,wang2024gft,xia2024opengraph} that can be shared among different graphs for GFM design. However, these methods primarily focus on graph-related tasks (e.g., node classification and link prediction), leaving the design of a GFM to enhance LLMs' reasoning ability unexplored.


\begin{wrapfigure}{r}{0.52\columnwidth}
    \vspace{-.5cm}
    \includegraphics[width=.52\columnwidth]{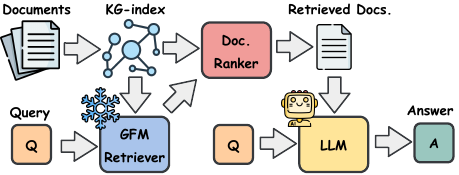}
    \vspace{-.5cm}
    \caption{The overview framework of \ourmethod. 
    }
    \label{fig:intro}
    \vspace{-0.3cm}
\end{wrapfigure}
To bridge the gap, in this paper, we propose an effective, efficient, and general graph foundation model for retrieval augmented generation (\ourmethod), thereby enhancing LLMs' reasoning ability.
As shown in \Cref{fig:intro}, we create a \emph{knowledge graph index} (KG-index) from documents in each dataset. The KG-index consists of interconnected factual triples pointing to the original documents, which serves as a structural knowledge index across multiple sources, enhancing the integration of diverse knowledge for complex reasoning tasks \citep{gutiérrez2024hipporag}. 
%
Then, we present the \emph{graph foundation model retriever} (GFM retriever), driven by a query-dependent GNN that captures complex query-knowledge relationships in a unified,  transferable space of semantics and graph structure. Through multi-layer message passing, the GFM retriever enables efficient multi-hop retrieval in a single step, surpassing previous multi-step methods.
The GFM retriever, with 8M parameters, undergoes a two-stage training: \emph{self-supervised KG completion pre-training} and \emph{supervised document retrieval fine-tuning} on large-scale datasets, including 60 knowledge graphs with over 14M triples and 700k documents. This large-scale training ensures the generalizability of GFM retriever to be applied to unseen datasets without further training.

In experiments, \ourmethod achieves state-of-the-art performance across three multi-hop QA datasets, demonstrating its effectiveness and efficiency in multi-hop reasoning. It also generalizes well across seven RAG datasets from diverse domains, such as biomedical, customer service, and general knowledge, without requiring additional training. Furthermore, \ourmethod follows the neural scaling law \citep{hestness2017deep}, whose performance benefits from training data and model size scaling, emphasizing its potential as a foundational model for future improvements.
The main contributions of this paper are as follows:
\begin{itemize}
\item We introduce a graph foundation model for retrieval augmented generation (\ourmethod), powered by a novel query-dependent GNN to enable efficient multi-hop retrieval within a single step.
\item We train a large-scale model with 8M parameters, marking the first graph foundation model (GFM) that can be applied directly to various unseen datasets for retrieval augmented generation.
\item We evaluate \ourmethod on three multi-hop QA datasets and seven domain-specific RAG datasets, achieving state-of-the-art performance across all, demonstrating its effectiveness, efficiency, generalizability, and potential as a foundational model for further enhancement.
\end{itemize}

\section{Related Work}
\noindent\textbf{Retrieval-augmented generation (RAG)} \citep{gao2023retrieval} provides an effective way to integrate external knowledge into large language models (LLMs) by retrieving relevant documents to facilitate LLM generation. Early works adopt the pre-trained dense embedding model to encode documents as separate vectors \citep{karpukhin2020dense,bge_m3,li2023towards,moreira2024nv}, which are then retrieved by calculating the similarity to the query. Despite efficiency and generalizability, these methods struggle to capture complex document relationships. Subsequent studies have explored multi-step retrieval, where LLMs guide an iterative process to retrieve and reason over multiple documents \citep{trivedi2023interleaving,jiang2023active,su-etal-2024-dragin}. However, this approach is computationally expensive. 

\noindent\textbf{Graph-enhanced retrieval augmented generation (GraphRAG)} \citep{peng2024graph,han2024retrieval} is a novel approach that builds graphs to explicitly model the complex relationships between knowledge, facilitating comprehensive retrieval and reasoning. Early research focuses on retrieving information from existing knowledge graphs (KGs), such as WikiData \citep{vrandevcic2014wikidata} and Freebase \citep{bollacker2008freebase}, by identifying relevant facts or reasoning paths \citep{li2023graph,luoreasoning,panda2024holmes}. Recent studies have integrated documents with KGs to improve knowledge coverage and retrieval \citep{edge2024local,liang2024kag}. A graph structure is built from these documents to aid in identifying relevant content for LLM generation \citep{dong2024don}. Based on graphs, LightRAG \citep{guo2024lightrag} incorporates graph structures into text indexing and retrieval, enabling efficient retrieval of entities and their relationships.
HippoRAG \citep{gutiérrez2024hipporag} enhances multi-hop retrieval by using a personalized PageRank algorithm to locate relevant knowledge with graphs. However, the graph structure can be noisy and incomplete, leading to suboptimal performance. Efforts to incorporate GNNs into graph-enhanced RAG \citep{mavromatis2024gnn,he2024g} have shown impressive results due to the multi-hop graph reasoning capabilities of GNNs in handling incomplete graphs \citep{yasunaga2021qa}. Nonetheless, these methods still limit in generalizability due to the lack of a graph foundational model.

\noindent\textbf{Graph Foundation models (GFM)} aims to be a large-scale model that can generalize to various datasets \citep{maoposition,liu2025graph}. The main challenge in designing GFMs is identifying graph tokens that capture invariance across diverse graph data. For instance, ULTRA \citep{galkintowards} employs four fundamental relational interactions in knowledge graphs (KGs) to create a GFM with 0.2M parameters for link prediction. OpenGraph \citep{xia2024opengraph} develops a graph tokenizer that converts graphs into a unified node token representation, enabling transformer-like GFMs for tasks such as link prediction and node classification. GFT \citep{wang2024gft} introduces a transferable tree vocabulary to construct a GFM that demonstrates effectiveness across various tasks and domains in graph learning. Despite these successful efforts, most methods primarily focus on conventional graph-related tasks, and transformer-like GFMs \citep{tang2024higpt,tang2024graphgpt} struggle with large-scale graphs and capture logical association~\citep{qiuunderstanding}. How to design a GNN-based GFM to enhance the reasoning of LLM remains an open question.


\section{Approach}\label{sec:approach}

The proposed \ourmethod essentially implements a GraphRAG paradigm by constructing graphs from documents and using a graph-enhanced retriever to retrieve relevant documents.

\begin{figure*}[t]
    \centering
    \includegraphics[width=1\textwidth]{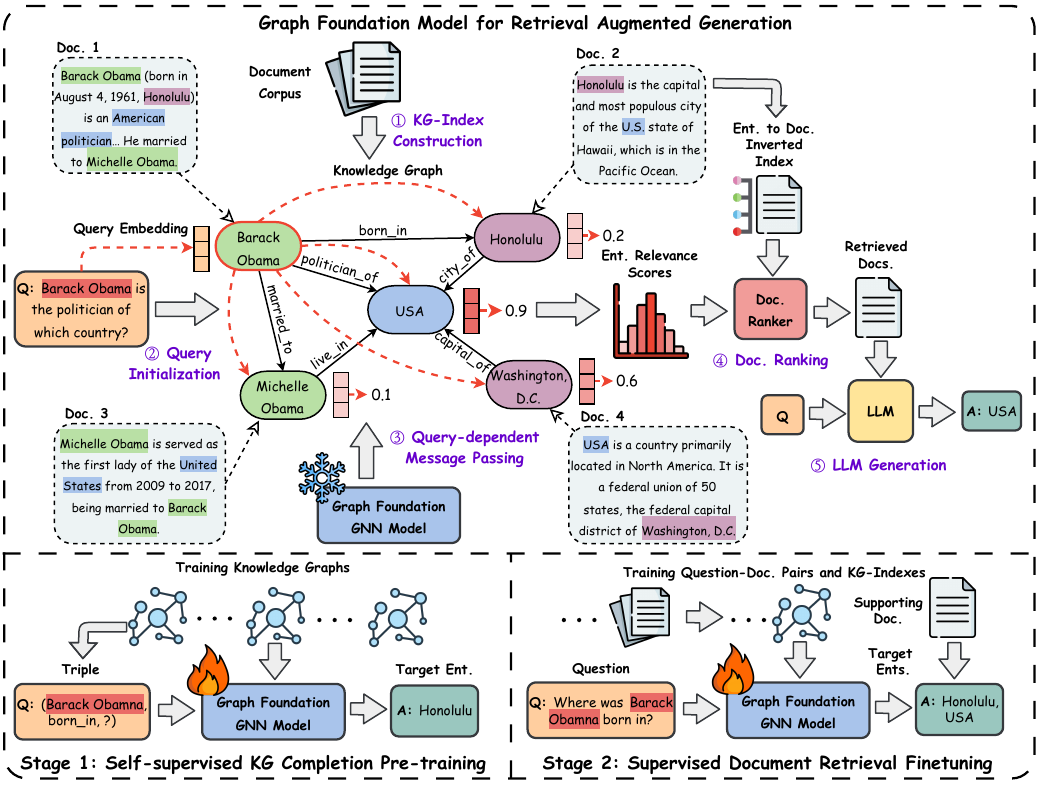}
    \vspace{-.3cm}
    \caption{The detailed framework of \ourmethod and training processes of graph foundation model. The \ourmethod consists of three main components: A. \emph{KG-index construction}, which constructs a knowledge graph index from document corpus (\tikzmarknode[mycircled,mypurple]{a1}{1}); B. \emph{graph foundation model retriever} (GFM retriever), which is pre-trained on large-scale datasets and could retrieve documents based on any user query and KG-index (\tikzmarknode[mycircled,mypurple]{a2}{2}\tikzmarknode[mycircled,mypurple]{a3}{3}); and C. \emph{documents ranking and answer generation}, which ranks retrieved documents and generates final answer (\tikzmarknode[mycircled,mypurple]{a4}{4}\tikzmarknode[mycircled,mypurple]{a5}{5}).
    }
    \label{fig:framework}
    \vspace{-0.5cm}
\end{figure*}

\noindent\textbf{GFM-RAG Overview.}
Given a set of documents $\gD=\{D_1,D_2,\ldots,D_{|\gD|}\}$, we construct a knowledge graph $\gG=\{(e,r,e')\in \gE\times\gR\times\gE\}$, where $e,e'\in\gE$ and $r\in\gR$ denote the set of entities and relations extracted from $\gD$, respectively. 
For a user query $q$, we aim to design a graph-enhanced retriever to obtain relevant documents from $\gD$ by leveraging the knowledge graph $\gG$. The whole \ourmethod process can be formulated as:
\begin{gather}
    \gG = \text{KG-index}(\gD), \\ \gD^K = \text{GFM-Retriever}(q,\gD,\gG),\\
    a = \text{LLM}(q,\gD^K).
\end{gather}
In the first step, KG-index($\cdot$)  constructs a knowledge graph index $\gG$ from the document corpus $\gD$, followed by our proposed \emph{graph foundation model retriever} (GFM-Retriever), which is pre-trained on large-scale datasets. It retrieves top-$K$ documents based on any user query $q$ and knowledge graph index $\gG$. The retrieved documents $\gD^K$, along with the query $q$, are then input into a large language model (LLM) to generate the final answer $a$.
%
These three main components in \ourmethod are illustrated in \Cref{fig:framework} and will be detailed next.

\subsection{KG-index Construction}\label{sec:kg-construction}
Conventional embedding-based index methods encode documents as separate vectors \citep{karpukhin2020dense,bge_m3,moreira2024nv}, which are limited in modeling the relationships between them. Knowledge graphs (KGs), on the other hand, explicitly capturing the relationships between millions of facts, can provide a structural index of knowledge across multiple documents \citep{edge2024local,gutiérrez2024hipporag}. The structural nature of the KG-index aligns well with the human hippocampal memory indexing theory \citep{teyler1986hippocampal}, where the KG-index functions like an artificial hippocampus to store associations between knowledge memories, enhancing the integration of diverse knowledge for complex reasoning tasks \citep{gutiérrez2024hipporag}.

To construct the KG-index, given a set of documents $\gD$, we first extract entities $\gE$ and relations $\gR$ to form triples $\gT$ from documents. Then, the entity to document inverted index $M \in \{0,1\}^{|\gE|\times|\gD|}$ is constructed to record the entities mentioned in each document. Such a process can be achieved by existing open information extraction (OpenIE) tools \citep{angeli2015leveraging,ijcai2022p793,pai2024survey}. To better capture the connection between knowledge, we further conduct the entity resolution \citep{gillick2019learning,zeakis2023pre} to add additional edges $\gT^{\texttt{+}}$ between entities with similar semantics, e.g., (\texttt{USA}, \texttt{equivalent}, \texttt{United States of America}). Therefore, the final KG-index $\gG$ is constructed as $\gG=\{(e,r,e')\in\gT\cup\gT^{\texttt{+}}\}$. In implementation, we leverage an LLM \citep{gpt4o} as the OpenIE tool (prompts are shown in \Cref{tab:prompt}) and a pre-trained dense embedding model \citep{santhanam2022colbertv2} for entity resolution. Details can be found in \Cref{app:kg_index}.

\subsection{Graph Foundation Model (GFM) Retriever}\label{sec:gfm-retriever}
The GFM retriever is designed to retrieve relevant documents based on any user query and the constructed KG-index. While the KG-index offers a structured representation of knowledge, it still suffers from incompleteness and noise, resulting in suboptimal retrieval performance when solely relying on its structure \citep{gutiérrez2024hipporag}. Recently, graph neural networks (GNNs) \citep{wu2020comprehensive} have shown impressive multi-hop reasoning ability by capturing the complex relationships between knowledge for retrieval or question answering \citep{mavromatis2024gnn,he2024g}. However, existing GNNs are limited in generalizability, as they are usually trained on specific graphs \citep{maoposition,liu2025graph}, which limits their application to unseen corpora and KGs. Therefore, there is still a need for a graph foundation model that can be directly applied to unseen datasets and KGs without additional training.

To address these issues, we propose the first graph foundation model-powered retriever (GFM retriever), which harnesses the graph reasoning ability of GNNs to capture the complex relationships between queries, documents, and knowledge graphs in a unified and transferable space. The GFM retriever employs a query-dependent GNN to identify relevant entities in graphs that will aid in locating pertinent documents. After pre-training on large-scale datasets, the GFM retriever can be directly applied to new corpora and KGs without further training.

\subsubsection{Query-dependent GNN}\label{sec:message-passing}
Conventional GNNs \citep{gilmer2017neural} follow the message passing paradigm, which iteratively aggregates information from neighbors to update entity representations. Such a paradigm is not suitable for the GFM retriever as it is graph-specific and neglects the relevance of queries. Recent query-dependent GNNs \citep{zhu2021neural,galkintowards} have shown promising results in capturing query-specific information and generalizability to unseen graphs, which is essential for the GFM retriever and can be formulated as: 
\begin{equation}
    \setlength\abovedisplayskip{2pt}
    \setlength\belowdisplayskip{2pt}
        H_q^L = \text{GNN}_q(q,\gG,H^0),
\end{equation}
where $H^0\in\sR^{|\gE|\times d}$ denotes initial entity features, and $H_q^L$ denotes the updated entity representations conditioned on query $q$ after $L$ layers of query-dependent message passing. 

The query-dependent GNN is theoretically proven to exhibit multi-hop logical reasoning ability \citep{huang2023theory,yasunaga2021qa,qiuunderstanding} (detailed in \Cref{app:theory}), which is selected as the backbone of our GFM retriever. It allows the GFM retriever to dynamically adjust the message passing process based on user queries and find the most relevant information on the graph with multi-hop reasoning. The path interpretation for this multi-hop reasoning process is shown in \Cref{sec:interpretation}.

\noindent\textbf{Query Initialization.} Given a query $q$, we first encode it into a query embedding with a sentence embedding model:
\begin{equation}
    \setlength\abovedisplayskip{2pt}
    \setlength\belowdisplayskip{2pt}
    \vq = \text{SentenceEmb}(q),~\vq\in\mathbb{R}^d,
\end{equation}
where $d$ denotes the dimension of the query embedding. Then, for all the entities mentioned in the query $e_q\in\gE_q\subseteq\gE$, we initialize their entity features as $\vq$ while others as zero vectors:
\begin{equation}
    \setlength\abovedisplayskip{1pt}
    \setlength\belowdisplayskip{1pt}
    H^0 = \begin{cases}
        \vq, & e\in\gE_q, \\
        \vzero, & \text{otherwise}.
    \end{cases}
\end{equation}

\noindent\textbf{Query-dependent Message Passing.} The query-dependent message passing will propagate the information from the question entities to other entities in the KG to capture their relevance to the query. The message passing process can be formulated as:
\begin{align}
    & \text{\textbf{Triple-level}: } \nonumber \\
    & h^0_r = \text{SentenceEmb}(r),~h^0_r\in\mathbb{R}^d, \\
    & m_e^{l+1} = \text{Msg}(h_e^l,g^{l+1}(h_r^l),h_{e'}^l), (e,r,e')\in\gG, \\
    & \text{\textbf{Entity-level}: } \nonumber \\
    & h_e^{l+1} = \text{Update}(h_e^l,\text{Agg}(\{m_{e'}^{l+1} | e'\in\gN_r(e),r\in\gR\})),
\end{align}
where $h^l_e, h^l_r$ denote the entity and relation embeddings at layer $l$, respectively. The relation embeddings $ h^0_r$ are also initialized using the same sentence embedding model as the query, reflecting their semantics (e.g., ``$\texttt{born\_in}$''), and updated by a layer-specific function $g^{l+1}(\cdot)$, implemented as a 2-layer MLP.
The $\text{Msg}(\cdot)$ is operated on all triples in the KG to generate messages, which is implemented with a non-parametric DistMult \citep{yang2015embedding} following the architecture of NBFNet \citep{zhu2021neural}. For each entity, we aggregate the messages from its neighbors $\gN_r(e)$ with relation $r$ using sum and update the entity representation with a single linear layer.

After $L$ layers message passing, a final MLP layer together with a sigmoid function maps the entity embeddings to their relevance scores to the query:
\begin{equation}
    \setlength\abovedisplayskip{2pt}
    \setlength\belowdisplayskip{2pt}
    P_q = \sigma(\text{MLP}(H_q^L)),~P_q\in\mathbb{R}^{|\gE|\times 1}.
\end{equation}

\noindent\textbf{Generalizability.} Since the query, entity, and relation embeddings are initialized using the same sentence embedding model with identical dimensions, the query-dependent GNN can be directly applied to different queries and KGs. This allows it to learn complex relationships between queries and entities by taking into account both the semantics and structure of the KG through training on large-scale datasets.

\subsubsection{Training Process}\label{sec:training}

\noindent\textbf{Training Objective.} The training objective of the GFM retriever is to maximize the likelihood of the relevant entities to the query, which can be optimized by minimizing the binary cross-entropy (BCE) loss:
\begin{equation}
    \setlength\abovedisplayskip{3pt}
    \setlength\belowdisplayskip{3pt}
    \gL_{\text{BCE}} = -\frac{1}{|\gA_q|}\sum_{e\in\gA_q} \log P_q(e) - \frac{1}{|\gE^{\texttt{-}}|}\sum_{|\gE^{\texttt{-}}|} \log (1-P_q(e)), 
\end{equation}
where $\gA_q$ denotes the set of target relevant entities to the query $q$, and $\gE^{\texttt{-}}\subseteq \gE\setminus \gA_q$ denotes the set of negative entities sampled from the KG. However, due to the sparsity of the target entities, the BCE loss may suffer from the gradient vanishing problem \citep{lin2024understanding}. To address this issue, we further introduce the ranking loss \citep{bai2023regression} to maximize the margin between the positive and negative entities:
\begin{equation}
    \setlength\abovedisplayskip{2pt}
    \setlength\belowdisplayskip{2pt}
    \gL_{\text{RANK}} = - \frac{1}{|\gA_q|}\sum_{e\in\gA_q} \frac{P_q(e)}{\sum_{e'\in\gE^{\texttt{-}}} P_q(e')}.
\end{equation}
The final training objective is the weighted combination of the BCE loss and ranking loss:
\begin{equation}
    \setlength\abovedisplayskip{2pt}
    \setlength\belowdisplayskip{2pt}
    \gL = \alpha\gL_{\text{BCE}} + (1-\alpha) \gL_{\text{RANK}}.\label{eq:training}
\end{equation}

\noindent\textbf{Self-supervised KG Completion Pre-training.} To enhance the graph reasoning capability of the GFM retriever, we first pre-train it on a large-scale knowledge graph (KG) completion task. We sample a set of triples from the KG index and mask either the head or tail entity to create synthetic queries in the form $q=(e,r,?)~\text{or}~(?, r, e')$, with the masked entity serving as the target entity $\gA_q = \{e\}~\text{or}~\{e'\}$. The GFM retriever is then trained to predict the masked entity using both the query and the KG, as outlined in \eqref{eq:training}.

\noindent\textbf{Supervised Document Retrieval Fine-tuning.} After self-supervised pre-training, we supervised fine-tune the GFM retriever on a labeled document retrieval task. In this task, queries $q$ are natural language questions, and target entities $\gA_q$ are extracted from labeled supporting documents $\gD_q$. The GFM retriever is trained to retrieve relevant entities from the KG index using the same training objective as in \eqref{eq:training}.

\subsection{Documents Ranking and Answer Generation}\label{sec:ranking}
Given the entity relevance scores $P_q\in\sR^{|\gE|\times 1}$ predicted by the GFM retriever, we first retrieve the top-$T$ entities $\gE_q^{T}$ with the highest relevance scores as:
\begin{equation}
    \gE_q^{T} = \arg\text{top-}T(P_q),~\gE_q^{T}=\{e_1,\ldots,e_T\}.\label{eq:top-T}
\end{equation}
These retrieved entities are then used by the document ranker to obtain the final documents. To diminish the influence of popular entities, we weight the entities by the inverse of their frequency as entities mentioned in the document inverted index $M \in \{0,1\}^{|\gE|\times|\gD|}$ and calculate the final document relevance scores by summing the weights of entity mentioned in documents:
{
\begin{gather}
    \setlength\abovedisplayskip{2pt}
    \setlength\belowdisplayskip{2pt}
    F_e = \begin{cases}
        \frac{1}{\sum_{d\in\gD} M[e,d]}, & e\in\gE_q^{T}, \\
        0, & \text{otherwise},
    \end{cases}\label{eq:entity_weight} \\
    P_d = M^{\top}F_e,~P_d\in\sR^{|\gD|\times 1}.\label{eq:doc_score}
\end{gather}
The top-$K$ documents are retrieved based on the document relevance scores $P_d$ and fed into the context of LLMs, with a retrieval augmented generation manner, to generate the final answer:
\begin{gather}
    \setlength\abovedisplayskip{1pt}
    \setlength\belowdisplayskip{1pt}
    \gD^K = \arg\text{top-}K(P_d),~\gD^K=\{D_1,\ldots,D_K\}, \\
    a = \text{LLM}(q,\gD^K).
\end{gather}
}
\section{Experiment}\label{sec:experiment}
In experiments, we aim to address the following research questions: (1) How does \ourmethod perform in multi-hop retrieval and QA tasks? (\Cref{sec:retrieval,sec:qa}); (2) What are the efficiency and effectiveness of \ourmethod in multi-hop retrieval? (\Cref{sec:efficiency}); (3) How well does \ourmethod generalize to unseen datasets as a foundation model? (\Cref{sec:generalizability}); (4) How does the performance of \ourmethod scale with training as a foundation model? (\Cref{sec:scaling}); (5) How to interpret \ourmethod in multi-hop reasoning? (\Cref{sec:interpretation}).

\subsection{Experimental Setup}\label{sec:setup}
\noindent\textbf{Datasets.} We first evaluate the effectiveness of \ourmethod on three widely-used multi-hop QA datasets, including HotpotQA \citep{yang2018hotpotqa}, MuSiQue \citep{trivedi2022musique}, and 2WikiMultiHopQA (2Wiki) \citep{ho2020constructing}. We also evaluate the performance of \ourmethod on seven RAG datasets from three domains, including biomedical \citep{jin-etal-2019-pubmedqa}, custom support \citep{sadat-etal-2023-delucionqa,nandy-etal-2021-question-answering,malaviya2023expertqa,castelli-etal-2020-techqa}, and general knowledge \citep{nguyen2016ms,kamalloo2023hagrid}, to demonstrate the generalizability of \ourmethod as the foundation model. The detailed statistics of the test datasets are shown in the \Cref{app:datasets}.

\noindent\textbf{Baselines.} We compare against several widely used retrieval methods under three categories: (1) \emph{single-step naive methods}: BM25 \citep{robertson1994some}, Contriever \citep{izacardunsupervised}, GTR \citep{ni2022large}, ColBERTv2 \citep{santhanam2022colbertv2}, RAPTOR \citep{sarthiraptor}, Proposition \citep{chen-etal-2024-dense}; (2) \emph{graph-enhanced methods}: GraphRAG (MS) \citep{edge2024local}, LightRAG \citep{guo2024lightrag}, HippoRAG \citep{gutiérrez2024hipporag}, SubgraphRAG~\citep{lisimple}, G-retriever~\citep{he2024g}; (3) \emph{multi-step methods}: Adaptive-RAG \citep{jeong2024adaptive}, FLARE \citep{jiang2023active}, and IRCoT \citep{trivedi2023interleaving} framework that can be integrated with arbitrary retrieval methods to conduct multi-step retrieval and reasoning. The detailed introduction of the baselines are shown in the \Cref{app:baselines}.

\noindent\textbf{Metrics.} For retrieval performance, we use recall@2 (R@2) and recall@5 (R@5) as evaluation metrics. For final QA performance, we use the EM score and F1 score following previous works \citep{gutiérrez2024hipporag}.

\noindent\textbf{Implementation Details.} 
The GFM retriever is implemented with 6 query-dependent message passing layers with the hidden dimension set to 512. The pre-trained all-mpnet-v2 \citep{all-mpnet-v2} is adopted as the sentence embedding model and is frozen during training. The total parameters of the GFM retriever are 8M, which is trained on 8 NVIDIA A100s (80G) with batch size 4, learning rate 5e-4, and loss weight $\alpha=0.3$. The training data contains 60 KGs with over 14M triples constructed from 700k documents extracted from the training set. 
The statistics of training data are shown in \Cref{tab:training_data}, and the implementations are detailed in \Cref{app:implementation}.

    \begin{table}[]
    \centering
    \caption{Retrieval performance comparison.}
    \label{tab:retrieval}
    \resizebox{.9\columnwidth}{!}{%
    \begin{tabular}{@{}c|l|cc|cc|cc@{}}
        \toprule
        \multicolumn{1}{c|}{\multirow{2}{*}{Category}} & \multicolumn{1}{c|}{\multirow{2}{*}{Method}} & \multicolumn{2}{c|}{HotpotQA} & \multicolumn{2}{c|}{MuSiQue} & \multicolumn{2}{c}{2Wiki}                                                 \\ \cmidrule(l){3-8}
        \multicolumn{1}{c|}{}                          & \multicolumn{1}{c|}{}                        & R@2                           & R@5                          & R@2                       & R@5           & R@2           & R@5           \\ \midrule
        \multirow{12}{*}{Single-step}                  & BM25                                         & 55.4                          & 72.2                         & 32.3                      & 41.2          & 51.8          & 61.9          \\
                                                       & Contriever                                   & 57.2                          & 75.5                         & 34.8                      & 46.6          & 46.6          & 57.5          \\
                                                       & GTR                                          & 59.4                          & 73.3                         & 37.4                      & 49.1          & 60.2          & 67.9          \\
                                                       & ColBERTv2                                    & 64.7                          & 79.3                         & 37.9                      & 49.2          & 59.2          & 68.2          \\
                                                       & RAPTOR                                       & 58.1                          & 71.2                         & 35.7                      & 45.3          & 46.3          & 53.8          \\
                                                       & Proposition                                  & 58.7                          & 71.1                         & 37.6                      & 49.3          & 56.4          & 63.1          \\\cmidrule{2-8}
                                                       & GraphRAG (MS)                                & 58.3
                                                       & 76.6
                                                       & 35.4
                                                       & 49.3
                                                       & 61.6
                                                       & 77.3

        \\
                                                       & LightRAG                                     & 38.8                          & 54.7                         & 24.8                      & 34.7          & 45.1          & 59.1          \\
                                                       & HippoRAG (Contriever)                        & 59.0                          & 76.2                         & 41.0                      & 52.1          & 71.5          & 89.5          \\
                                                       & HippoRAG (ColBERTv2)                         & 60.5                          & 77.7                         & 40.9                      & 51.9          & 70.7          & 89.1          \\ 
                                                       & SubgraphRAG                                  & 61.5                          & 73.0                         & 42.1                      & 49.3          & 70.7          & 85.5          \\
                                                       & G-retriever                                   & 53.3                         & 65.5                         & 38.8                     & 45.1        & 60.8         & 67.8         \\

\midrule
        \multirow{7}{*}{Multi-step}                                                                           & Adaptive-RAG                                 & 61.0                          & 76.4                         & 35.1                      & 44.7          & 44.7          & 61.4          \\
                                                       & FLARE                                        & 73.1                          & 81.3                         & 44.3                      & 55.1          & 67.1          & 73.1          \\
        & IRCoT + BM25                                 & 65.6                          & 79.0                         & 34.2                      & 44.7          & 61.2          & 75.6          \\
                                                       & IRCoT + Contriever                           & 65.9                          & 81.6                         & 39.1                      & 52.2          & 51.6          & 63.8          \\
                                                       & IRCoT + ColBERTv2                            & 67.9                          & 82.0                         & 41.7                      & 53.7          & 64.1          & 74.4          \\
                                                       & IRCoT + HippoRAG (Contriever)                & 65.8                          & 82.3                         & 43.9                      & 56.6          & 75.3          & 93.4          \\
                                                       & IRCoT + HippoRAG (ColBERTv2)                 & 67.0                          & 83.0                         & 45.3                      & 57.6          & 75.8          & 93.9          \\
        \midrule
        Single-step                                    & \ourmethod                                   & \textbf{78.3}                 & \textbf{87.1}                & \textbf{49.1}             & \textbf{58.2} & \textbf{90.8} & \textbf{95.6} \\ \bottomrule
    \end{tabular}%
    }
\end{table}
\subsection{Retrieval Performance}\label{sec:retrieval}
We first evaluate the retrieval performance of \ourmethod against the baselines on three multi-hop QA datasets. As shown in \Cref{tab:retrieval}, \ourmethod achieves the best performance on all datasets, outperforming the SOTA IRCoT + HippoRAG by 16.8\%, 8.3\%, 19.8\% in R@2 on HotpotQA, MuSiQue, and 2Wiki, respectively. The results demonstrate the effectiveness of \ourmethod in multi-hop retrieval. From the result, we can observe that the naive single-step retrievers (e.g., BM25, RAPTOR) are outperformed by graph-enhanced HippoRAG, which highlights the significance of graph structure in multi-hop retrieval. Although GraphRAG (MS) and LightRAG use the graph structure, it struggles with multi-hop QA tasks as its retriever is designed for summarization and lacks multi-hop reasoning capability.
With the help of LLMs, the multi-step retrieval pipeline IRCoT improves the performance of all single-step methods through iterative reasoning and retrieval. However, \ourmethod still outperforms the multi-step methods by a large margin even with a single-step retrieval. This indicates that the \ourmethod can effectively conduct the multi-hop reasoning in a single step (detailed in \Cref{sec:interpretation} and \Cref{app:multi_hop}), which is more efficient and effective than the multi-step retrieval pipeline (detailed in \Cref{sec:efficiency}).

\subsection{Question Answering Performance}\label{sec:qa}
\begin{table}
    \centering
    \caption{Question answering performance comparison.}
    \label{tab:qa}
    \resizebox{.9\columnwidth}{!}{%
    \begin{tabular}{@{}c|l|cc|cc|cc@{}}
    \toprule
    \multirow{2}{*}{Category}    & \multicolumn{1}{c|}{\multirow{2}{*}{Retriever}} & \multicolumn{2}{c|}{HotpotQA} & \multicolumn{2}{c|}{MuSiQue} & \multicolumn{2}{c}{2Wiki} \\ \cmidrule(l){3-8} 
                                 & \multicolumn{1}{c|}{}                           & EM            & F1            & EM            & F1           & EM          & F1          \\ \midrule
    \multirow{5}{*}{Single-step} & None                                            & 30.4          & 42.8          & 12.5          & 24.1         & 31.0        & 39.0        \\
                                 & ColBERTv2                                       & 43.4          & 57.7          & 15.5          & 26.4         & 33.4        & 43.3        \\
                                    & GraphRAG (MS) & 35.3 & 54.6 & 13.4 & 29.5 & 28.3 & 46.9 \\
                                 & LightRAG & 36.8 & 48.3 & 18.1 & 27.5 & 45.1 & 49.5 \\
                                 & HippoRAG (ColBERTv2)                            & 41.8          & 55.0          & 19.2          & 29.8         & 46.6        & 59.5        \\ \midrule
    \multirow{4}{*}{Multi-step}                                   & Adaptive-RAG & 45.5	& 59.6	& 13.8	& 25.6	& 48.9	& 62.8 \\
                                 & FLARE & 48.7 & 60.6 & 16.2 & 28.4 & 46.7 & 65.4 \\ 
    & IRCoT (ColBERTv2)                               & 45.5          & 58.4          & 19.1          & 30.5         & 35.4        & 45.1        \\
                                 & IRCoT + HippoRAG (ColBERTv2)                    & 45.7          & 59.2          & 21.9          & 33.3         & 47.7        & 62.7        \\ 
                                 \midrule
                                 Single-step                  & \ourmethod                                      & {\ul 51.6}    & {\ul 66.9}    & {\ul 30.2}    & {\ul 40.4}    & {\ul 69.8}    & {\ul 77.7}    \\
    Multi-step                   & IRCoT + \ourmethod                              & \textbf{56.0} & \textbf{71.8} & \textbf{36.6} & \textbf{49.2} & \textbf{72.5} & \textbf{80.8}       \\ \bottomrule
    \end{tabular}%
    }
   \end{table}
We then evaluate the QA performance of \ourmethod, as it is directly influenced by retrieval quality. We adopt the GPT-4o-mini \citep{gpt4o} as LLM and use the top-$5$ retrieved documents for generating answers. From the results shown in \Cref{tab:qa}, the single-step \ourmethod has already achieved state-of-the-art performance against all other baselines. Meanwhile, we also integrate \ourmethod with IRCoT to conduct multi-step retrieval and reasoning, which further improves the performance by 8.5\%, 21.2\%, 3.9\% in EM on three datasets, respectively. The results demonstrate the effectiveness and great compatibility of \ourmethod with an arbitrary multi-step framework in multi-hop reasoning tasks.

\subsection{Efficiency Analysis}\label{sec:efficiency}
\begin{table}
    \centering
    \caption{Retrieval efficiency and performance comparison.}
    \label{tab:efficiency}
    \resizebox{.8\columnwidth}{!}{%
        \begin{tabular}{@{}l|cc|cc|cc@{}}
            \toprule
            \multirow{2}{*}{Method} & \multicolumn{2}{c|}{HotpotQA} & \multicolumn{2}{c|}{MuSiQue} & \multicolumn{2}{c}{2Wiki}                                                  \\
                                    & Time (s)                      & R@5                          & Time (s)                  & R@5           & Time (s)       & R@5           \\ \midrule
            ColBERTv2               & \textbf{0.035}                & 79.3                         & \textbf{0.030}            & 49.2          & \textbf{0.029} & 68.2          \\
            HippoRAG                & 0.255                         & 77.7                         & 0.251                     & 51.9          & 0.158          & 89.1          \\
            LightRAG                & 0.861                         & 54.7                         & 1.109                     & 34.7          & 0.911          & 59.1          \\
            GraphRAG (MS)          & 2.759                         & 76.6                         & 3.037                     & 49.3          & 1.204          & 77.3          \\ \midrule
            IRCoT + ColBERTv2       & 1.146                         & 82.0                         & 1.152                     & 53.7          & 2.095          & 74.4          \\
            IRCoT + HippoRAG        & 3.162                         & 83.0                         & 3.104                     & 57.6          & 3.441          & 93.9          \\ \midrule
            \ourmethod              & {\ul 0.107}                   & \textbf{87.1}                & {\ul 0.124}               & \textbf{58.2} & {\ul 0.060 }   & \textbf{95.6} \\ \bottomrule
        \end{tabular}%
    }
\end{table}
\ourmethod achieves great efficiency in performing multi-step reasoning in a single step. As shown in \Cref{tab:efficiency}, while the naive single-step methods get the best efficiency whose performance is not satisfying. Admittedly, the multi-step framework IRCoT could improve the performance, but it suffers from high computational costs due to the iterative retrieval and reasoning with LLMs. In contrast, \ourmethod conducts multi-hop reasoning within a single-step GNN reasoning, which is more effective than single-step methods and more efficient than multi-step ones.

\subsection{Ablation Study}\label{sec:ablation}
We conduct ablation studies to investigate the effectiveness of different components in \ourmethod, including: different sentence embedding models (\Cref{app:text_embeddings}), pre-training strategies (\Cref{app:pre-training}), loss weighting strategies (\Cref{app:loss_weight}), ranking methods (\Cref{app:ranking_method}), training datasets (\Cref{app:training_data_ablation}), and the construction of KG-index (\Cref{app:cost}). The results show that \ourmethod is not sensitive to different sentence embedding models, and the pre-training strategy, as well as the loss weighting strategy, are both crucial for the performance of \ourmethod.

\subsection{Model Generalizability}\label{sec:generalizability}
\begin{figure}[t]
    \centering
    \begin{minipage}[t]{0.48\linewidth}
        \centering
        \includegraphics[width=1\linewidth]{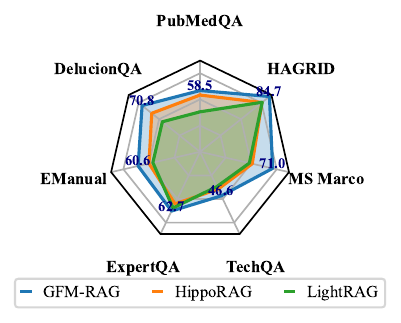}
        \caption{Model generalizability comparison.}
        \label{fig:radar}
    \end{minipage}%
    \hfill
    \begin{minipage}[t]{0.48\linewidth}
        \centering
        \includegraphics[width=1\linewidth]{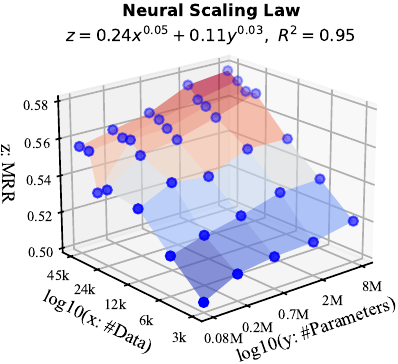}
        \caption{Neural scaling law of \ourmethod.}
        \label{fig:scaling}
    \end{minipage}
\end{figure}

To demonstrate the generalizability of \ourmethod as a foundation model, we test the performance (R@5) of \ourmethod on seven RAG datasets without any domain-specific fine-tuning. Specifically, we first build the KG-index from the documents in each dataset. Then, given the query, we use the pre-trained GFM retriever to retrieve the top-$K$ documents with the help of the corresponding KG-index.
As shown in \Cref{fig:radar}, \ourmethod achieves the best performance on all datasets, outperforming the SOTA HippoRAG by 18.9\% on average. The results demonstrate the generalizability of \ourmethod as the foundation model which can be directly applied to various unseen datasets without any domain-specific fine-tuning. Additionally, results in \Cref{app:trans} demonstrate \ourmethod's strong transferability for further performance improvement when fine-tuned on domain-specific datasets.

\begin{table*}[]
    \centering
    \caption{Path interpretations of GFM for multi-hop reasoning, where $r^{-1}$ denotes the inverse of original relation.}
    \label{tab:cases}
    \resizebox{.9\textwidth}{!}{%
    \begin{tabular}{@{}c|p{5in}@{}}
        \toprule
        \textbf{Question}             &  What \textit{football club} was owned by the singer of "\textit{Grow Some Funk of Your Own}"?
        \\ \midrule
        \textbf{Answer}               & Watford Football Club
        \\ \midrule
        \textbf{Sup. Doc.}        & [ ``Grow Some Funk of Your Own'', ``Elton John''] \\\midrule
        \textbf{Paths}                                   & \begin{tabular}[c]{@{}p{5in}@{}} 1.095: (grow some funk of your own, is a song by, elton john) $\to$ (elton john, equivalent, sir elton hercules john) $\to$ (sir elton hercules john, named a stand $\text{after}^{-1}$, \textbf{watford football club})   \\
        0.915: (grow some funk of your own, is a song by, elton john) $\to$ (elton john, equivalent, sir elton hercules john) $\to$  (sir elton hercules john, owned, \textbf{watford football club})
        \end{tabular} \\ \midrule\midrule
        \textbf{Question}             &  When was the judge born who made notable contributions to the trial of the man who tortured, raped, and murdered eight student nurses from \textit{South Chicago Community Hospital} on the night of \textit{July 13-14, 1966}? 
        \\ \midrule
        \textbf{Answer}               & June 4, 1931
        \\ \midrule
        \textbf{Sup. Doc.}        & [ ``Louis B. Garippo'', ``Richard Speck''] \\\midrule
        \textbf{Paths}                                   & \begin{tabular}[c]{@{}p{5in}@{}} 
            0.797: (south chicago community hospital, committed crimes at$^{-1}$, richard speck) $\to$ (richard speck, equivalent, trial of richard speck) $\to$ (trial of richard speck, made contributions during$^{-1}$, \textbf{louis b  garippo})
            \\
            0.412: (south chicago community hospital, were from$^{-1}$, eight student nurses) $\to$ (eight student nurses, were from, south chicago community hospital) $\to$ (south chicago community hospital, committed crimes at$^-1$, \textbf{richard speck})
        \end{tabular} \\
        \bottomrule
    \end{tabular}
    }
\end{table*}

\subsection{Model Neural Scaling Law}\label{sec:scaling}
We further investigate the neural scaling law of \ourmethod, which quantifies how model performance grows with the scale of training data and model parameter size. It has been validated in the recent foundation models \citep{kaplan2020scaling,dehghani2023scaling}. As shown in \Cref{fig:scaling}, the performance of \ourmethod (MRR: $z$) scales well with the training data ($x$) and the model size ($y$), which can be fitted by the power-law scaling law $z \propto 0.24x^{0.05} + 0.11y^{0.03}$. The results demonstrate the scalability of \ourmethod as the foundation model and potential for further improvement. The detailed analysis of the neural scaling law is shown in \Cref{app:scaling}.

\subsection{Path Interpretations}\label{sec:interpretation}
\ourmethod exhibits the multi-hop reasoning ability powered by the multi-layer GFM. We provide path interpretations of \ourmethod for multi-hop reasoning in \Cref{tab:cases}. Inspired by NBFNet \citep{zhu2021neural}, the paths' importance to the final prediction can be quantified by the partial derivative of the prediction score with respect to the triples at each layer (hop), defined as:
\begin{equation}
    \setlength\abovedisplayskip{0pt}
    \setlength\belowdisplayskip{0pt}
    s_1,s_2,\ldots,s_L=\arg\mathop{\text{top-}}k\frac{\partial p_e(q)}{\partial s_l}.
\end{equation}
The top-$k$ path interpretations can be obtained by the top-$k$ longest paths with beam search. We illustrate the path interpretations in \Cref{tab:cases}. In the first example, \ourmethod successfully deduces that the singer of the song has a football club named after him and that he owned it. In the second example, \ourmethod identifies two paths related to the murder case and the judge presiding over the trial. These interpretations show that \ourmethod exhibits the ability of multi-hop reasoning within single-step retrieval. We also illustrate the distribution the multi-hop prediction in \Cref{app:multi_hop}.
\section{Conclusion}\label{sec:conclusion}
In this paper, we introduce the first graph foundation model for retrieval augmented generation. By leveraging the knowledge graph index, \ourmethod explicitly models the complex relationships between knowledge and documents, facilitating a more effective and efficient retrieval process. Powered by a query-dependent GNN pre-trained on large-scale datasets, \ourmethod can effectively perform multi-hop reasoning over the graph structure to find relevant knowledge in a single step. Extensive experiments across three benchmark datasets and seven domain-specific datasets demonstrate that \ourmethod significantly outperforms state-of-the-art methods in effectiveness, efficiency, and generalizability. Its alignment with scaling laws also suggests the potential for scaling to even larger datasets. In the future, we plan to conduct larger-scale training and further explore \ourmethod's capabilities in other challenging scenarios such as knowledge graph completion and question answering.

\clearpage
\section*{Acknowledgments}
G Haffari is partly supported by the ARC Future Fellowship FT190100039 and DARPA Assured Neuro Symbolic Learning and Reasoning (ANSR) program under award number FA8750-23-2-1016. C Gong is supported by NSF of China (Nos: 62336003, 12371510). D Phung is supported by the Australian Research Council (ARC) Discovery Project DP250100262 and DP230101176. S Pan was partly funded by Australian Research Council (ARC) under grants FT210100097 and DP240101547 and the CSIRO – National Science Foundation (US) AI Research Collaboration Program. 
\bibliographystyle{plainnat}
\bibliography{sections/main}







\appendix

\addcontentsline{toc}{section}{Appendix} 
\part{Appendix} 
\parttoc 
\section{Query-dependent GNNs for Multi-hop Reasoning and Retrieval}\label{app:theory}
We provide a detailed explanation about why query-dependent GNNs can be used for multi-hop reasoning and retrieval. Recent studies \citep{huang2023theory,qiuunderstanding} have theoretically proven that query-dependent GNNs are effective for capturing the multi-hop logical associations on KGs to answer queries, such as:
\begin{equation}
    \exists y: \texttt{politician\_of}(\text{Barack Obama},y) \leftarrow \texttt{work\_in}(\text{Barack Obama},z_1) \wedge \texttt{city\_of}(z_1, y),
\end{equation}
where the right part denotes the logical associations can be executed to answer the query on the left, i.e., ``\texttt{politician\_of}(\text{Barack Obama},y)''.

This query is semantic equivalent to the nature language question: ``\texttt{Barack Obama is the politician of which country?}''. By treating the input question as a ``soft query'' (query in nature language), we apply the query-dependent GNN (GFM) to bridge the gap between nature language and logical query. The GFM tries to understand the semantic of the questions and learn to conduct complex logical reasoning (e.g., multi-hop reasoning) on KGs for retrieval~\citep{yasunaga2021qa}. The learned logical associations for reasoning are shown in \Cref{sec:interpretation}.

\begin{table}[h]
    \centering
    \caption{Statistics of the query-doc pairs and KGs used for training.}
    \label{tab:training_data}
        \begin{tabular}{@{}ccccccc@{}}
            \toprule
            Dataset & \#Q-doc Pair & \#Document & \#KG & \#Entity & \#Relation & \#Triple  \\ \midrule
            HotpotQA & 20,000        & 204,822    & 20    & 1,930,362  & 967,218     & 6,393,342  \\
            MuSiQue  & 20,000        & 410,380    & 20    & 1,544,966  & 900,338     & 4,848,715  \\
            2Wiki    & 20,000        & 122,108    & 20    & 916,907    & 372,554     & 2,883,006  \\
            \midrule
            \textbf{Total} & \textbf{60,000} & \textbf{737,310} & \textbf{60} & \textbf{4,392,235} & \textbf{2,240,110} & \textbf{14,125,063} \\ \bottomrule
        \end{tabular}%
\end{table}
\begin{table}[]
\centering
\caption{Statistics of the datasets and constructed KG-indexes used for testing.}
\label{tab:test_data}
\resizebox{.9\columnwidth}{!}{%
\begin{tabular}{@{}ccccccc@{}}
\toprule
Dataset    & Domain            & \#Test & \#Document & \#Entity & \#Relation & \#Triple \\ \midrule
HotpotQA   & Multi-hop         & 1,000 & 9,221     & 87,768  & 45,112    & 279,112 \\
MuSiQue    & Multi-hop         & 1,000 & 11,656    & 100,853 & 55,944    & 319,618 \\
2Wiki      & Multi-hop         & 1,000 & 6,119     & 48,779  & 20,748    & 160,950 \\
PubMedQA   & Biomedical        & 2,450 & 5,932     & 42,389  & 20,952    & 149,782 \\
DelucionQA & Customer Support  & 184   & 235       & 2,669   & 2,298     & 6,183   \\
TechQA     & Customer Support  & 314   & 769       & 10,221  & 4,606     & 57,613  \\
ExpertQA   & Customer Support  & 203   & 808       & 11,079  & 6,810     & 16,541  \\
EManual    & Customer Support  & 132   & 102       & 695     & 586       & 1,329   \\
MS Marco   & General Knowledge & 423   & 3,481     & 24,740  & 17,042    & 63,995  \\
HAGRID     & General Knowledge & 1,318 & 1,975     & 23,484  & 18,653    & 48,969  \\ \bottomrule
\end{tabular}%
}
\end{table}
\section{Datasets}\label{app:datasets}
\subsection{Multi-hop QA Datasets}\label{app:multi_hop_qa}
Three multi-hop QA datasets are used in our experiments: HotpotQA \citep{yang2018hotpotqa}, MuSiQue \citep{trivedi2022musique}, and 2WikiMultiHopQA (2Wiki) \citep{ho2020constructing}. We provide a brief overview of these datasets below.
\begin{itemize}
    \item HotpotQA \citep{yang2018hotpotqa} is a multi-hop QA dataset that requires reasoning over multiple documents to answer questions. The dataset consists of 97k question-answer pairs, where each question is associated with up to 2 supporting and several distracting documents. The questions are designed to be answerable using multiple pieces of information from the supporting documents.
    \item MuSiQue \citep{trivedi2022musique} is a challenging multi-hop QA dataset with 25k 2-4 hop questions. It requires coherent multi-step reasoning to answer questions that span multiple documents. 
    \item 2WikiMultiHopQA (2Wiki) \citep{ho2020constructing} is a multi-hop QA dataset that requires reasoning over multiple Wikipedia articles to answer questions. The dataset consists of 192k questions, which are designed to be answerable using information from 2 or 4 articles.
\end{itemize}
In experiments, we adhere to the official data split to obtain the training samples and follow existing methods \citep{trivedi2023interleaving,gutiérrez2024hipporag} to use the same 1,000 samples from each validation set to avoid data leakage. We merge the candidate passages as the document corpus for KG-index construction. The statistics of the training and test data are presented in \Cref{tab:training_data} and \Cref{tab:test_data}, respectively.

\subsection{Domain-specific RAG Datasets}\label{app:domain_specific}
To test the generalizability of \ourmethod, we evaluate it on seven domain-specific RAG datasets \citep{friel2024ragbench} including, (1) \emph{biomedical}: PubMedQA \citep{jin-etal-2019-pubmedqa}; (2) \emph{customer support}: DelucionQA \citep{sadat-etal-2023-delucionqa}, TechQA \citep{castelli-etal-2020-techqa}, ExpertQA \citep{malaviya2023expertqa}, EManual \citep{nandy-etal-2021-question-answering}; (3) \emph{general knowledge}: MS Marco \citep{nguyen2016ms}, HAGRID \citep{kamalloo2023hagrid}. We provide a brief overview of these datasets below.

\begin{itemize}
    \item PubMedQA \citep{jin-etal-2019-pubmedqa} is a collection of PubMed research abstracts with corresponding questions paired with 4 abstract chunks.
    \item  DelucionQA \citep{sadat-etal-2023-delucionqa} is a domain-specific RAG dataset leveraging Jeep’s 2023 Gladiator model manual as the source of knowledge, where each question is associated with 4 context documents and only 1 relevant passage.
    \item TechQA \citep{castelli-etal-2020-techqa} is a collection of real-world user questions posted on IBMDeveloper and DeveloperWorks forums, along with 10 technical support documents relating to each question.
    \item ExpertQA \citep{malaviya2023expertqa} is a collection of curated questions from domain experts in various fields of science, arts, and law. The dataset also contains expert-curated passages relevant to each question.
    \item EManual \citep{nandy-etal-2021-question-answering} is a question-answering dataset comprising consumer electronic device manuals and realistic questions about them composed by human annotators, where each question is related with up to 3 context documents.
     \item MS Marco \citep{nguyen2016ms} is an open-domain question-answering dataset sourced from Bing search engine user query logs. Each question is associated with 10 context passages retrieved via Bing web search.
    \item HAGRID \citep{kamalloo2023hagrid} is a multi-lingual information retrieval dataset with questions and passages from MIRACL \citep{zhang2022making}.
\end{itemize}
In experiments, we use test sets constructed by RAGBench \citep{friel2024ragbench} and merge all the candidate passages as document corpus for KG-index construction. The statistics of the test dataset are detailed in \Cref{tab:test_data}.

\section{Baselines}\label{app:baselines}
In experiments, we compare with several widely used retrieval methods under three categories: (1) \emph{single-step naive methods}: BM25 \citep{robertson1994some}, Contriever \citep{izacardunsupervised}, GTR \citep{ni2022large}, ColBERTv2 \citep{santhanam2022colbertv2}, RAPTOR \citep{sarthiraptor}, Proposition \citep{chen-etal-2024-dense}; (2) \emph{graph-enhanced methods}: GraphRAG (MS) \citep{edge2024local}, LightRAG \citep{guo2024lightrag}, HippoRAG \citep{gutiérrez2024hipporag}; (3) \emph{multi-step methods}: Adaptive-RAG \citep{jeong2024adaptive}, FLARE \citep{jiang2023active}, and IRCoT \citep{trivedi2023interleaving}. The detailed introduction of the baselines is as follows.

\noindent\textbf{Single-step Naive Methods} are widely adopted in real-world applications due to their great efficiency and generalizability. 
\begin{itemize}
    \item BM25 \citep{robertson1994some} is a classic information retrieval method based on the probabilistic model that ranks a set of documents based on the query terms frequency appearing in each document.
    \item Contriever \citep{izacardunsupervised} trains a dense retriever with contrastive learning on a large-scale corpus to retrieve relevant documents for a given query.
    \item GTR \citep{ni2022large} develops a scale-up T5-based dense retriever that could generalize across different datasets and domains.
    \item ColBERTv2 \citep{santhanam2022colbertv2} is a state-of-the-art dense retriever that couples an aggressive residual compression mechanism with a denoised supervision strategy to simultaneously improve the retrieval quality.
    \item RAPTOR \citep{sarthiraptor} is an LLM-augmented retriever that recursively embeds, clusters, and summarizes chunks of text, constructing a tree with differing levels of summarization to enable accurate retrieval.
    \item Proposition \citep{chen-etal-2024-dense} enhances the performance of dense retrievers by leveraging LLMs to generate a natural language proposition that captures the essential information of the document.
\end{itemize}

\noindent\textbf{Graph-enhanced Methods} design a retriever that is built upon a graph structure to conduct effective retrieval and reasoning.
\begin{itemize}
    \item GraphRAG (MS) \citep{edge2024local} is a graph-enhanced retrieval method originally proposed by Microsoft. It builds a graph structure from the document corpus and use hierarchical community detection to cluster the documents into communities and generate a summary for each community. The summary together with the original documents are retrieved by the retriever for LLM generation. 
    \item LightRAG \citep{guo2024lightrag} is an innovative graph-enhanced RAG method that incorporates graph structures into text indexing and retrieval, enabling efficient retrieval of entities and their relationships. It employs a dual-level retrieval system to gather both low-level and high-level knowledge for LLM generation.
    \item HippoRAG \citep{gutiérrez2024hipporag} is a state-of-the-art, training-free graph-enhanced retriever that uses the Personalized PageRank algorithm to assess entity relevance to a query and performs multi-hop retrieval on a document-based knowledge graph. It can be directly applied to various datasets.
\end{itemize}

\noindent\textbf{Multi-step Methods} are designed to conduct multi-hop reasoning by iteratively retrieving and reasoning over documents, which can be integrated with arbitrary retrieval methods.
\begin{itemize}
    \item Adaptive-RAG \citep{jeong2024adaptive} proposes an adaptive multi-step retrieval method that can dynamically select the most suitable retrieval strategy based on the complexity of the query.
    \item FLARE \citep{jiang2023active} introduces a multi-step retrieval method that actively decide when and how to retrieve documents. It also predicts the future content to the guide the retrieval in next steps. 
    \item IRCoT \citep{trivedi2023interleaving} is a powerful multi-step retrieval pipeline that integrates the retrieval with the chain-of-thought (CoT) reasoning of LLMs. It guides the retrieval with CoT and in turn using retrieved documents to improve CoT. IRCoT can be compatible with arbitrary retrievers to conduct multi-step retrieval and reasoning.
\end{itemize}

\section{Implementations and Training Details}\label{app:implementation}

\subsection{Training Data Construction}\label{app:kg_index}
We extract 60,000 samples from the training set of HotpotQA, MuSiQu, and 2Wiki to construct KG-indexes and conduct large-scale training. Specifically, we merge the candidate passages as the document corpus. In the KG-index construction, we use the GPT-4o-mini \citep{gpt4o} with the OpenIE prompts described in HippoRAG \citep{gutiérrez2024hipporag} to extract the entities, relations, and triples from the document corpus. Then, we use the ColBERTv2 \citep{santhanam2022colbertv2} to conduct the entity resolution by computing the similarity between entities as
\begin{equation}
    s(e_i, e_j) = \text{Emb.}(e_i)^{\top} \text{Emb.}(e_j), 
\end{equation}
where a new triple $(e_i, \texttt{equivalent}, e_j)$ is generated if $s(e_i, e_j) > \tau$ and $e_i \neq e_j$. We set the threshold $\tau$ as 0.8 in our experiments. We divide the samples into groups of approximately 1k questions and 10k documents each to control the constructed KG-index size. In the end, we obtain 60 different KG-indexes and associated question-document pairs for model training.

\subsection{Model Settings}\label{app:model_settings}
In \ourmethod, the GFM is implemented as a 6-layer query-dependent GNN with the hidden dimension of 512, DistMult message function, and sum aggregation. The relation update function $g^{l}(\cdot)$ is implemented as a 2-layer MLP. We use the all-mpnet-v2 as the sentence embedding model with a dimension of 768. The total training parameters of the GFM is 8M. In the retrieval stage, we select top $T=20$ entities for the document ranker.

\begin{table}[]
\centering
\caption{The detailed implementation and training settings of \ourmethod.}
\label{tab:settings}
\begin{tabular}{@{}ccc@{}}
\toprule
                                       & Setting                  & \ourmethod           \\ \midrule
\multirow{3}{*}{KG-index Construction} & OpenIE                   & GPT-4o-mini          \\
                                       & Entity resolution        & ColBERTv2            \\
                                       & $\tau$                   & 0.8                  \\ \midrule
\multirow{7}{*}{GFM Model}             & \# Layer                  & 6                    \\
                                       & Hidden dim               & 512                  \\
                                       & Message                  & DistMult             \\
                                       & Aggregation              & Sum                  \\
                                       & $g^{l}(\cdot)$           & 2-layer MLP          \\
                                       & Sentence embedding model & all-mpnet-v2         \\ 
                                       & Doc. ranker entities $T$ & 20                   \\\midrule
\multirow{6}{*}{KGC Pre-training}                                             & $\alpha$            & 1             \\
& Optimizer               & AdamW                \\
                                       & Learning rate            & 5e-4             \\
                                       & Batch size               & 4                    \\
                                       & Training steps           & 30,000               \\
                                       & \# Negative sample        & 128                  \\ \midrule
\multirow{6}{*}{Supervised Retrieval Fine-tuning}                                        & $\alpha$            & 0.3             \\
& Optimizer               & AdamW                \\
                                       & Learning rate            & 5e-4             \\
                                       & Batch size               & 4                    \\
                                       & Training epochs          & 5                    \\
                                       & \# Negative sample        & $\gE\setminus \gA_q$ \\ \bottomrule
\end{tabular}%
\end{table}

\subsection{Training Settings}\label{app:training_settings}
In KG completion pre-training, we randomly sample triples $(e, r, t)$ from knowledge graphs and mask out either the head or the tail entity to create a synthetic query $q = (e, r, ?)$ and answer $a = \{e\}$ in a self-supervised manner. For example, given a triple (\texttt{Barack Obama}, \texttt{born\_in}, \texttt{Honolulu}), we can create a query as $(\texttt{Barack Obama}, \texttt{born\_in}, ?)$, which is encoded as a sentence embedding and fed into the GFM to predict the target entity \texttt{Honolulu} on graphs.

In supervised document retrieval fine-tuning, we obtain natural language questions and supporting documents from the multi-hop QA datasets. For each question, we identify the entities from its supporting documents as the targets. For instance, given the question \textit{``Where was Barack Obama born in?''}, we can extract two entities such as \texttt{[Honolulu, USA]} from its supporting documents (e.g., Doc.~2 in \Cref{fig:framework}). The GFM is trained to maximize the likelihood of these two target entities.

In the self-supervised KG completion pre-training, the GFM is trained on the mixture of 60 constructed KG-indexes for 30,000 steps. Then, we conduct the supervised document retrieval fine-tuning on the labeled question-document pairs for 5 epochs. The weight $\alpha$ between losses is set to 0.3. We use AdamW optimizer, learning rate of 5e-4 with batch sizes of both training stages set to 4. Each batch contains only one KG-index and training samples associated to it, where we randomly sample from different KG-indexes during training. The model is trained on 8 NVIDIA A100s (80G) with 14 hours pre-training and 5 hours supervised fine-tuning. The detailed settings are summarized in \Cref{tab:settings}.

\section{Additional Experiments}\label{app:additional_experiments}

\begin{table}[]
    \centering
    \caption{Comparison of different sentence embedding models used in \ourmethod.}
    \label{tab:text-emb}
    \begin{tabular}{@{}l|cc|cc|cc@{}}
    \toprule
    \multirow{2}{*}{Sentence Embedding Model} & \multicolumn{2}{c|}{HotpotQA} & \multicolumn{2}{c|}{MuSique}  & \multicolumn{2}{c}{2Wiki}     \\ \cmidrule(l){2-7} 
                                              & R@2           & R@5           & R@2           & R@5           & R@2           & R@5           \\ \midrule
    sentence-transformers/all-mpnet-base-v2   & \textbf{70.2} & \textbf{82.1} & 46.0 & 55.1 & \textbf{81.1} & 85.6 \\
    BAAI/bge-large-en                         & 68.1          & 81.1          & 45.9          &  \textbf{55.9}          & 80.7          & \textbf{86.3}          \\
    Alibaba-NLP/gte-Qwen2-1.5B-instruct       & 69.9          & 81.5          & 46.0          & 55.0          & 79.8          & 86.2          \\
    Alibaba-NLP/gte-Qwen2-7B-instruct         & 68.5          & 81.5          & 45.5          & 55.1          & 80.8          & 85.6          \\
    nvidia/NV-Embed-v2                        & 69.2          & 81.4          & \textbf{46.3}          & 54.9          & 80.3          & 85.5          \\ \bottomrule
    \end{tabular}%
    \end{table}

    \begin{table}[]
        \centering
        \caption{Comparison of \ourmethod with pre-trained and fine-tuned sentence embedding models.}
        \label{tab:text_emb_ablation}
        \begin{tabular}{@{}l|cc|cc|cc@{}}
        \toprule
        \multirow{2}{*}{Method}    & \multicolumn{2}{c|}{HotpotQA} & \multicolumn{2}{c|}{MuSiQue}  & \multicolumn{2}{c}{2Wiki}     \\ \cmidrule(l){2-7} 
                                   & R@2           & R@5           & R@2           & R@5           & R@2           & R@5           \\ \midrule
        \ourmethod                    & \textbf{78.3} & \textbf{87.1} & \textbf{49.1} & \textbf{58.2} & \textbf{90.8} & \textbf{95.6} \\ \midrule
        all-mpnet-v2 (pre-trained) & 59.4          & 73.3          & 33.2          & 46.3          & 48.5          & 59.4          \\
        all-mpnet-v2 (finetuned)   & 67.0          & 82.3          & 41.7          & 55.0          & 65.1          & 76.7          \\ \bottomrule
        \end{tabular}%
        \end{table}
\subsection{Effectiveness of Different Sentence Embeddings}\label{app:text_embeddings}

In this section, we first study the effectiveness of different sentence embeddings in the GFM. We compare the all-mpnet-v2 \citep{all-mpnet-v2}, bge-large-en \citep{bge_embedding}, gte-Qwen2-1.5B-instruct and gte-Qwen2-7B-instruct \citep{li2023towards} as well as NV-Embed-v2 \citep{lee2024nv}. We download the official pre-trained model from the Huggingface\footnote{\url{https://huggingface.co/}}. The details of the models are shown in \Cref{tab:text-emb}. From the results, we can  observe that the performance variance between different sentence embeddings is relatively small, where the all-mpnet-v2 achieves the best performance with respect to 3 metrics. This indicates that \ourmethod is not sensitive to the choice of sentence embedding models. In experiments, we use the all-mpnet-v2 as the default sentence embedding model due to its efficiency. However, it has relative smaller context-size (512) which limits the length of input text. We leave the exploration of larger context-size sentence embedding models (e.g., NV-Embed-v2 with 32k context) for future work.

Then, we expand our ablation study to compare \ourmethod with variants without GNN and using solely the pre-trained all-mpnet-v2 embeddings and those fine-tuned on multi-hop QA data, respectively. The results are shown in \Cref{tab:text_emb_ablation}. We can observe that GNN plays a crucial role in retrieval. The sentence embedding model all-mpnet-v2 is pre-trained on large-scale text data and could potentially see the QA data. However, it is not specifically trained for the multi-hop QA task, which leads to suboptimal performance in capturing the relationship between question and supporting documents. The fine-tuned all-mpnet-v2 achieves better performance than the pre-trained one by supervised fine-tuning on the multi-hop QA data, but still inferior to \ourmethod. This indicates that the GNN can effectively capture the relationship between knowledge and conduct multi-hop reasoning, which is not achievable by simply using the sentence embedding model.

\subsection{Effectiveness of Different Training Strategies}\label{app:pre-training}
In this section, we first study the effectiveness of the two training tasks used in \ourmethod. We compare the performance by only conducting the KG completion pre-training (\ourmethod $w/o$ Fine-tune) and supervised document retrieval fine-tuning (\ourmethod $w/o$ Pre-train). The results are shown in \Cref{tab:pre-train}. The results show that removing the supervised document retrieval fine-tuning significantly decreases the performance of \ourmethod. This highlights the importance of supervised fine-tuning, as it enables the model to understand users' queries and better capture the relevance between questions and knowledge for improved retrieval. 

\begin{table}[]
    \centering
    \caption{Effectiveness of KGC pre-training and supervised retrieval fine-tuning in \ourmethod.}
    \label{tab:pre-train}
    \begin{tabular}{@{}l|cc|cc|cc@{}}
    \toprule
    \multirow{2}{*}{Method} & \multicolumn{2}{c|}{HotpotQA} & \multicolumn{2}{c|}{MuSique}  & \multicolumn{2}{c}{2Wiki}     \\ \cmidrule(l){2-7} 
                                              & R@2           & R@5           & R@2           & R@5           & R@2           & R@5           \\ \midrule
    \ourmethod                                & \textbf{78.3} & \textbf{87.1} & \textbf{49.1} & 58.2          & \textbf{89.1} & \textbf{92.8} \\ 
    \ourmethod $w/o$ Retrieval Fine-tune                &   21.0            &  32.8             &     18.3          &   25.9            &  44.6             &   53.4            \\
    \ourmethod $w/o$ KGC Pre-train                & 77.8          & 86.5          & 48.3          & \textbf{58.3} & 88.3          & 92.5          \\\bottomrule
    \end{tabular}%
    \end{table}

    \begin{table}[]
        \centering
        \caption{Knowledge graph completion result of different training strategies.}
        \label{tab:kgc}
        \begin{tabular}{@{}l|cccc@{}}
        \toprule
        Method                                                              & MRR            & Hits@1         & Hits@3         & Hits@10        \\ \midrule
        \ourmethod                                                            & 0.193          & 0.138          & 0.221          & 0.293          \\
        \ourmethod $w/o$ Retrieval Fine-tune                                            & \textbf{0.304} & \textbf{0.234} & \textbf{0.323} & \textbf{0.451} \\
        \begin{tabular}[c]{@{}l@{}}\ourmethod $w/o$ KGC Pre-train\end{tabular} & 0.029          & 0.007          & 0.022         & 0.067           \\ \bottomrule
        \end{tabular}%
        \end{table}

Although the pre-training has a relatively small impact on the final performance, its primary purpose is to learn the general graph reasoning ability, following previous studies like ULTRA \cite{galkintowards}. This would enhance the generalization and robustness of the GFM, which could be beneficial to its performance on other tasks, such as knowledge graph completion. To further validate this, we conduct an ablation study to compare \ourmethod with different training strategies on the knowledge graph completion task. We report the knowledge graph completion (KGC) performance on the KG-index from the test set of the HotpotQA dataset. The results are shown in \Cref{tab:kgc}. 

From the knowledge graph completion results, we can observe that the \ourmethod undergoes only the pre-training (\ourmethod $w/o$ Fine-tune) achieves the best performance, which indicates that the pre-training is effective in learning the general graph reasoning ability. The performance of \ourmethod with only supervised fine-tuning (\ourmethod $w/o$ Pre-train) is significantly lower than that of \ourmethod with pre-training. This indicates that the supervised fine-tuning is only learning the specific downstream task, which would limit the generalization ability of \ourmethod as the foundation model. The GFM trained with both pre-training and supervised fine-tuning achieves the second-best performance on the knowledge graph completion task and the best performance on the multi-hop QA task. This indicates that both training strategies are essential for \ourmethod to learn the general graph reasoning ability and benefit specific downstream tasks.

\subsection{Effectiveness of Loss Weights}\label{app:loss_weight}
In this section, we examine the effectiveness of the weights assigned to the BCE loss and ranking loss in training \ourmethod. We compare performance by varying the weight $\alpha$ between the two losses: $\gL=\alpha\gL_{\text{BCE}} + (1-\alpha)\gL_{\text{RANK}}$, with results presented in \Cref{tab:loss_weight}. The findings indicate that using only either the BCE loss or ranking loss leads to suboptimal performance ($\alpha=0~\text{or}~1$). The best performance occurs when $\alpha$ is set to 0.3, which aligns with previous studies \citep{lin2024understanding} suggesting that a smaller weight for BCE loss is preferable when positive samples are rare in the training data.

\begin{table}[]
    \centering
    \caption{Effectiveness (MRR) for the weight $\alpha$ of two losses.}
    \label{tab:loss_weight}
    \begin{tabular}{@{}c|ccc@{}}
        \toprule
        $\alpha$ & HotpotQA        & MuSique         & 2Wiki           \\ \midrule
        0        & 0.5189          & 0.3252          & 0.4425          \\
        1        & 0.5096          & 0.3214          & 0.4282          \\
        0.7      & 0.5202          & 0.3249          & 0.4348          \\
        0.3      & \textbf{0.5243} & \textbf{0.3260} & \textbf{0.4490} \\ \bottomrule
        \end{tabular}%
    \end{table}

\subsection{Effectiveness of Ranking Methods}\label{app:ranking_method}
In this section, we investigate the effectiveness of different ranking methods based on inverted index used in \ourmethod. We compare four ranking methods including (1) \emph{IDF + Top-T Pred}: Our proposed method (\cref{eq:top-T,eq:entity_weight,eq:doc_score}), which maps the top-T entities predicted by GFM to documents using inverse document frequency (IDF)-weighted scores. (2) \emph{IDF + All Pred}: Uses all predicted entities from GFM and weights them by IDF ($w/o$ \cref{eq:top-T}). (3) \emph{Top-T Pred}: Uses only the top-T predicted entities without applying IDF weighting ($w/o$ \cref{eq:entity_weight}). (4) \emph{All Pred}: Use all entity predictions and directly map to document scores ($w/o$ \cref{eq:top-T,eq:entity_weight}). The results are shown in \Cref{tab:ranking}. The results show that the proposed \emph{IDF + Top-k Pred} performs the best. This indicates that the inverted index is a crucial component of \ourmethod, which serves as a bridge between structured reasoning over KGs and the unstructured documents required by LLMs, necessitating a careful design.

We acknowledge the potential alternatives, and as a promising future direction, we plan to explore end-to-end models that can jointly reason over structured and unstructured knowledge without relying on an explicit inverted index.

\begin{table}[h]
\centering
\caption{Comparison of different ranking methods.}
\label{tab:ranking}
\begin{tabular}{@{}lcccccc@{}}
\toprule
\multirow{2}{*}{Ranking Method} & \multicolumn{2}{c}{HotpotQA} & \multicolumn{2}{c}{MuSiQue} & \multicolumn{2}{c}{2Wiki} \\ \cmidrule(l){2-7} 
                                & R@2           & R@5          & R@2          & R@5          & R@2         & R@5         \\ \midrule
IDF + Top-T Pred (\ourmethod)     & \textbf{78.3} & \textbf{87.1} & \textbf{49.1} & \textbf{58.2} & \textbf{90.8} & \textbf{95.6} \\ \midrule
IDF + All Pred ($w/o$ \cref{eq:top-T})                 & 68.1          & 71.4         & 35.8         & 41.2         & 86.0        & 87.5        \\
Top-T Pred ($w/o$ \cref{eq:entity_weight})                     & 71.6          & 78.6         & 46.3         & 52.5         & 74.7        & 78.1        \\
All Pred ($w/o$ \cref{eq:top-T,eq:entity_weight})                      & 77.6          & 82.9         & 41.1         & 46.9         & 88.6        & 90.4        \\ \bottomrule
\end{tabular}%
\end{table}

\subsection{Ablation Study of Training Datasets}\label{app:training_data_ablation}
We further conducted ablation studies where \ourmethod is trained separately on each dataset, and we report performance across all three benchmarks. Results are shown \Cref{tab:dataset_ablation}. These results show that \ourmethod not only performs well on the trained datasets, but also generalizes well to other datasets.  More importantly, the model trained on multi-domain datasets performs competitively across all datasets, validating its ability to generalize effectively across domains and benefit from training on diverse KGs by learning generalizable reasoning ability across domains.

\begin{table}[]
\centering
\caption{Ablation study of \ourmethod trained on each dataset. Best results are highlighted in \textbf{bold}. The second best is \underline{underlined}.}
\label{tab:dataset_ablation}
\begin{tabular}{@{}c|cc|cc|cc@{}}
\toprule
Test Dataset     & \multicolumn{2}{c|}{HotpotQA} & \multicolumn{2}{c|}{MuSiQue}  & \multicolumn{2}{c}{2Wiki}     \\ \midrule
Training Dataset & R@2           & R@5           & R@2           & R@5           & R@2           & R@5           \\ \midrule
HotpotQA         & \textbf{79.3} & \textbf{87.8} & 46.9          & 57.2          & 86.6          & 92.4          \\
MusiQue          & 68.8          & 81.8          & {\ul 47.6}    & {\ul 57.5}    & 84.4          & 89.6          \\
2Wiki            & 72.2          & 77.9          & 46.6          & 55.5          & {\ul 89.3}    & {\ul 93.2}    \\ \midrule
All              & {\ul 78.3}    & {\ul 87.1}    & \textbf{49.1} & \textbf{58.2} & \textbf{90.8} & \textbf{95.6} \\ \bottomrule
\end{tabular}%
\end{table}

\subsection{Model Transferability}\label{app:trans}
\begin{table}[]
\centering
\caption{Model performance (R@5) and transferability comparsion.}
\label{tab:trans}
\resizebox{\columnwidth}{!}{%
\begin{tabular}{@{}lcccccc@{}}
\toprule
Model                  & \multicolumn{1}{l}{DelucionQA} & \multicolumn{1}{l}{EManual} & \multicolumn{1}{l}{ExpertQA} & \multicolumn{1}{l}{TechQA} & \multicolumn{1}{l}{MS Marco} & \multicolumn{1}{l}{HAGRID} \\ \midrule
HippoRAG (zero-shot)   & 59.0                           & 50.0                        & 55.1                         & 39.5                       & 51.1                         & 75.5                       \\
LightRAG (zero-shot)   & 46.1                           & 46.2                        & 59.4                         & 36.8                       & 48.3                         & 75.9                       \\
\ourmethod (zero-shot) & 70.8                           & 60.6                        & 62.7                         & 46.6                       & 71.0                         & 84.7                       \\
\ourmethod (domain-specific fine-tuning) & \textbf{82.7}                  & \textbf{75.9}               & \textbf{60.8}                & \textbf{49.5}              & \textbf{77.5}                & \textbf{86.6}              \\ \bottomrule
\end{tabular}%
}
\end{table}
In this section, we evaluate \ourmethod's transferability by conducting domain-specific fine-tuning on the training split of dataset on each domain. As shown in \ref{tab:trans}, \ourmethod performs well in zero-shot generalization, with further improvements achieved through domain-specific fine-tuning. This highlights its transferability when adapted to domain-specific datasets.

\begin{table}[]
    \centering
    \caption{The hidden dimension with corresponding model size and training batch size for scaling law analysis.}
    \label{tab:model_size}
    \begin{tabular}{@{}ccc@{}}
    \toprule
    Hidden Dim. & Parameter Size & Batch size (A100, 80G) \\ \midrule
    32          & 78,977         & 40                     \\
    64          & 215,297        & 20                     \\
    128         & 659,969        & 20                     \\
    256         & 2,237,441      & 8                      \\
    512         & 8,144,897      & 4                      \\ \bottomrule
    \end{tabular}%
    \end{table}
\subsection{Details of Model Neural Scaling}\label{app:scaling}
\begin{figure*}[]
    \centering
    \includegraphics[width=5.5in]{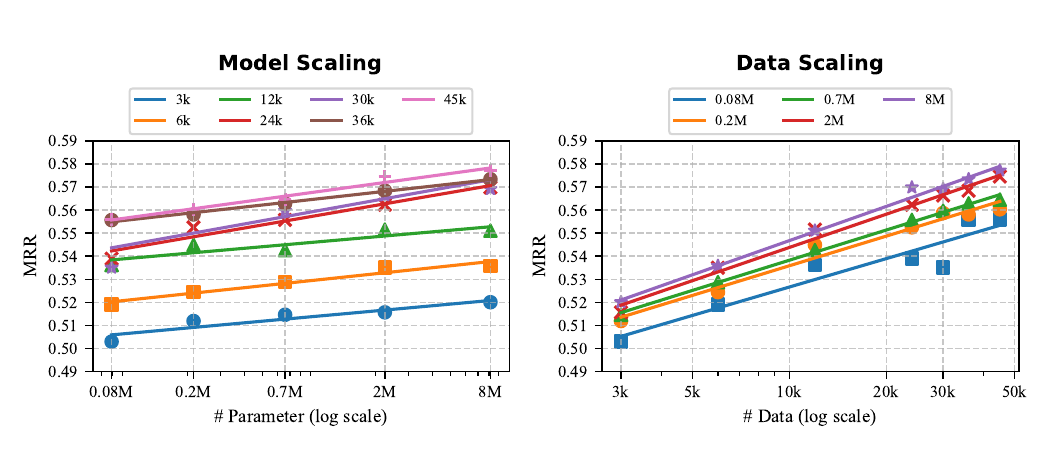}
    \caption{The illustration of the model and data scaling law of \ourmethod.}
    \label{fig:scaling_separate}
\end{figure*}

In this section, we provide more details on the neural scaling experiments. We evaluate the changes of the model performance with respect to different parameter sizes and training data sizes. In \ourmethod, the model parameter sizes are primarily influenced by the hidden dimension of the GFM. Thus, we vary the dimension from 32 to 512 which results in the model parameter sizes ranging from 0.08M to 8M. The detailed settings are shown in \Cref{tab:model_size}. We test models with different sizes on different scales of training data ranging from 3k to 45k samples. We separately report the fitted trend line of performance changing with model parameter size and training data size in \Cref{fig:scaling_separate}. From the trend line, we can observe that the performance of \ourmethod increases with the model parameter size and training data size. Meanwhile, with the larger model parameter size a larger training data size is required to achieve the best performance. This indicates that the performance of \ourmethod can be further improved by scaling up the model size and training data simultaneously.

To further investigate architectural design, we varied the number of GNN layers from 1 to 8 while keeping the hidden dimension fixed (512), and evaluated model performance across all datasets. The results are shown in \Cref{tab:layers}. We observe that performance generally improves with deeper GNN layers, which we attribute to both the increased model sizes and the ability to capture more complex multi-hop associations. This trend aligns with the neural scaling laws observed in foundation models, where larger parameter counts typically yield better generalization.

Interestingly, we find that performance peaks around 4 layers in some cases. As discussed in \Cref{app:theory} and \Cref{sec:interpretation}, \ourmethod is designed to capture logical associations from KGs through multi-hop message passing. However, since the maximum number of reasoning hops required by our datasets is 4, additional layers beyond this offer limited benefit, likely due to the absence of higher-hop training signals. This finding supports our hypothesis that \ourmethod effectively learns query-relevant multi-hop reasoning paths, and that deeper architectures may not improve performance without datasets requiring more complex reasoning.
In summary, these results demonstrate the effectiveness and interpretability of the proposed GNN-based architecture, and confirm that both model capacity and logical expressibility contribute to \ourmethod’s strong performance. We recognize the potential of other architectural designs and aim to explore them in the future, inspiring the community to do the same.

    \begin{table}[]
\centering
\caption{The different number of layers with corresponding model size and performance for scaling law analysis.}
\label{tab:layers}
\begin{tabular}{@{}c|cc|cc|cc|cc@{}}
    \toprule
Hidden Dim. = 512     & \multicolumn{2}{c|}{Averge}   & \multicolumn{2}{c|}{HotpotQA} & \multicolumn{2}{c|}{MuSiQue}  & \multicolumn{2}{c}{2Wiki}     \\ \midrule
L-Layer       & R@2           & R@5           & R@2           & R@5           & R@2           & R@5           & R@2           & R@5           \\ \midrule
1-layer (3M)  & 53.9          & 66.7          & 59.3          & 74.2          & 40.7          & 50.2          & 61.8          & 75.7          \\
2-layer (4M)  & 69.9          & 78.6          & 73.6          & 85.4          & 47.6          & 57.0          & 88.6          & 93.3          \\
4-layer (6M)  & 72.2          & \textbf{80.1} & 78.4          & 87.8          & 49.3          & \textbf{60.1} & 88.8          & 92.5          \\
6-layer (8M)  & 71.9          & 79.6          & 78.0          & 87.0          & 48.4          & 58.7          & 89.3          & 93.1          \\
8-layer (10M) & \textbf{73.0} & 79.9          & \textbf{79.7} & \textbf{87.8} & \textbf{49.7} & 59.1          & \textbf{89.5} & \textbf{92.8}\\ \bottomrule
\end{tabular}%
\end{table}

\begin{figure*}[t]
    \centering
    \includegraphics[width=5.5in]{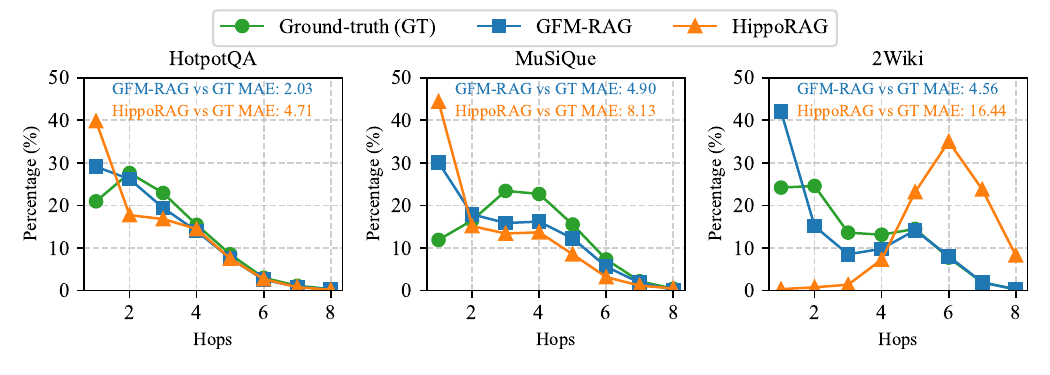}
    \caption{Statistics of the prediction hops of \ourmethod and HippoRAG against the ground-truth.}
    \label{fig:vis_hops}
\end{figure*}
\subsection{Visualization of the Distribution of Multi-hop Prediction}\label{app:multi_hop}
In this section, we visualize the distribution of the number of hops in the multi-hop reasoning process of \ourmethod. We calculate the number of hops in the ground-truth reasoning path required for each question in the test set of HotpotQA, MuSiQue, and 2Wiki. Then, we compare the distribution of the number of hops in the reasoning path of the ground-truth and the predicted reasoning path by \ourmethod as well as HippoRAG.
The results are shown in \Cref{fig:vis_hops}. We can observe that the distribution of \ourmethod is closely aligned to the ground-truth, which indicates that \ourmethod can effectively conduct the multi-hop reasoning within a single step. Meanwhile, the distribution of HippoRAG is relatively different from the ground-truth, especially in 2Wiki dataset. This indicates that HippoRAG may not be able to effectively capture the complex relationship to conduct multi-hop reasoning on graphs.

\begin{table}[]
\centering
\caption{The cost of the KG-index construction.}
\label{tab:kg_index_cost}
\begin{tabular}{@{}ccc@{}}
\toprule
LLM         & Price per 10k docs. & Total Price \\ \midrule
GPT-4o-mini & \$2.93                 & \$216        \\ \bottomrule
\end{tabular}%
\end{table}

\begin{table}[]
\centering
\caption{Token cost comparison for index construction}
\label{tab:token_cost}
\begin{tabular}{@{}cc@{}}
\toprule
Method     & \# Tokens per 10k documents \\
\midrule
LightRAG   & 55M                        \\
GraphRAG   & 76M                        \\
\ourmethod & \textbf{48M}              \\ \bottomrule
\end{tabular}%
\end{table}

\begin{table}[]
\centering
\caption{Graph Indexing time comparison.}
\label{tab:index_time}
\begin{tabular}{@{}cc@{}}
\toprule
Method        & Indexing time (s) \\ \midrule
LightRAG      & 1430.32           \\
GraphRAG (MS) & 1796.43           \\
\ourmethod       & \textbf{93.55}    \\ \bottomrule
\end{tabular}%
\end{table}

\begin{table}[]
\centering
\caption{Comparison of the model performance under the KG-index constructed by different LLMs. }
\label{tab:kg_index_llm}
\begin{tabular}{@{}l|cc|cc|cc@{}}
\toprule
\multirow{2}{*}{Method}  & \multicolumn{2}{c|}{HotpotQA} & \multicolumn{2}{c|}{MuSiQue} & \multicolumn{2}{c}{2Wiki} \\ \cmidrule(l){2-7} 
                         & R@2           & R@5           & R@2           & R@5          & R@2         & R@5         \\ \midrule
GFM-RAG (gpt-4o-mini)    & 78.3          & 87.1          & 49.1          & 58.2         & 90.8        & 95.6        \\
HippoRAG (gpt-4o-mini)   & 62.2          & 79.3          & 41.7          & 53.6         & 72.1        & 89.5        \\ \midrule
GFM-RAG (gpt-3.5-trubo)  & 75.6          & 84.7          & 46.1          & 55.8         & 85.2        & 90.4        \\
HippoRAG (gpt-3.5-trubo) & 60.5          & 77.7          & 40.9          & 51.9         & 70.7        & 89.1        \\ \bottomrule
\end{tabular}%
\end{table}
\subsection{Cost and Impact of LLMs on KG-index Construction}\label{app:cost}

In this section, we first analyze the cost of the KG-index construction. In experiments, we utilize GPT-4o-mini\footnote{\url{https://platform.openai.com/docs/models/o4-mini}} for OpenIE extraction and construct the KG-index for 737,310 documents. The cost is shown in \Cref{tab:kg_index_cost,tab:token_cost}. Specifically, we find that constructing the KG-index requires approximately 48M tokens per 10k documents, which costs around \$2.6 using GPT-4o-mini. LightRAG and GraphRAG cost 57M tokens and 76M tokens, respectively. Compared to other methods, \ourmethod is more cost-effective as it does not require generating community-level summaries. In addition, we compare the graph index construction time of \ourmethod in \Cref{tab:index_time}. Results show that \ourmethod benefits from the quick index process during retrieval, as it does not construct a traditional vector database to store documents and entities.

Admittedly, using an LLM for KG index construction incurs computational costs. However, KG construction has been extensively studied, and numerous alternative methods exist that do not rely on LLMs \citep{zhong2023comprehensive}. Our implementation offers an easy interface for integration with any KG construction tools. We would explore the use of other KG construction methods in future work.

We further analyze the impact of LLMs used for KG-index construction on the performance of \ourmethod. We conduct experiments using different LLMs for KG-index construction, including GPT-4o-mini and GPT-3.5-turbo\footnote{\url{https://platform.openai.com/docs/models/gpt-3-5-turbo}}. Then, we reevaluate the performance of \ourmethod and HippoRAG with the constructed KG-index. The results are shown in \Cref{tab:kg_index_llm}. From the results, the performance of both methods on the KG extracted by GPT-4o-mini is higher than the ones by GPT-3.5-turbo. This supports the opinion that GPT-4o-mini generally outperforms GPT-3.5-turbo in constructing high quality KG-index, which is crucial for the graph-enhanced retrieval. However, the performance of \ourmethod is significantly higher than HippoRAG under both KG-indexes. This indicates that \ourmethod is more robust to the quality of the KG-index, demonstrating the effectiveness of the GFM in graph reasoning and retrieval. 

\section{Prompts}\label{app:prompt}

In experiments, we follow the prompts used in HippoRAG \citep{gutiérrez2024hipporag} to extract the triples from the document corpus, which is shown in \Cref{tab:prompt}.

\begin{table}
    \centering
    \begin{tcolorbox}[title=OpenIE Prompt]
    \small
    \begin{lstlisting}
## Instruction:  
Your task is to construct an RDF (Resource Description Framework) graph from the given passages and named entity lists. Respond with a JSON list of triples, with each triple representing a relationship in the RDF graph. Pay attention to the following requirements: 
- Each triple should contain at least one, but preferably two, of the named entities in the list for each passage.
- Clearly resolve pronouns to their specific names to maintain clarity.  

Convert the paragraph into a JSON dict, it has a named entity list and a triple list.

## One-Shot Demonstration:
Paragraph:
```
Radio City
Radio City is India's first private FM radio station and was started on 3 July 2001. It plays Hindi, English and regional songs. Radio City recently forayed into New Media in May 2008 with the launch of a music portal - PlanetRadiocity.com that offers music related news, videos, songs, and other music-related features.
```
{
    "named_entities":
    ["Radio City", "India", "3 July 2001", "Hindi", "English", "May 2008", "PlanetRadiocity.com"]
}
{
    "triples": [
            ["Radio City", "located in", "India"],
            ["Radio City", "is", "private FM radio station"],
            ["Radio City", "started on", "3 July 2001"],
            ["Radio City", "plays songs in", "Hindi"],
            ["Radio City", "plays songs in", "English"]
            ["Radio City", "forayed into", "New Media"],
            ["Radio City", "launched", "PlanetRadiocity.com"],
            ["PlanetRadiocity.com", "launched in", "May 2008"],
            ["PlanetRadiocity.com", "is", "music portal"],
            ["PlanetRadiocity.com", "offers", "news"],
            ["PlanetRadiocity.com", "offers", "videos"],
            ["PlanetRadiocity.com", "offers", "songs"]
    ]
}

## Input
Convert the paragraph into a JSON dict, it has a named entity list and a triple list. Paragraph:
```
INPUT PASSAGE
```
        \end{lstlisting}
    \end{tcolorbox}
    \caption{The prompt for OpenIE extraction.}
\label{tab:prompt}
\end{table}

\section{Limitations}\label{app:Limitation}

The limitations of \ourmethod are as follows: (1) The construction of KG-index can be costly and time-consuming, especially when using LLMs for OpenIE extraction. We would explore the use of efficient KG construction methods in future work and optimize the construction process. (2) The model size of the \ourmethod is relatively small (8M) compared to other foundation models like large language models with billions of parameters. Although it is not faired to directly compare the GNN-based model with transformer-based LLMs, we would explore the scaling of \ourmethod in future work to improve its performance and generalizability. (3) \ourmethod is only evaluated on multi-hop QA tasks and KG completion tasks. We would explore the capabilities of \ourmethod in other tasks such as knowledge graph question answering and knowledge graph reasoning in future work to validate its effectiveness as a foundation model.

\section*{Impact Statement}
This paper presents work whose goal is to advance the field of Machine Learning. There are many potential societal consequences of our work, none which we feel must be specifically highlighted here.

\newpage
\section*{NeurIPS Paper Checklist}

\begin{enumerate}

\item {\bf Claims}
    \item[] Question: Do the main claims made in the abstract and introduction accurately reflect the paper's contributions and scope?
    \item[] Answer: \answerYes{} 
    \item[] Justification: The main claims in the abstract and introduction accurately reflect the paper's contributions and scope.

\item {\bf Limitations}
    \item[] Question: Does the paper discuss the limitations of the work performed by the authors?
    \item[] Answer: \answerYes{} 
    \item[] Justification: We have discussed the limitations of the work in \Cref{app:Limitation}.

\item {\bf Theory assumptions and proofs}
    \item[] Question: For each theoretical result, does the paper provide the full set of assumptions and a complete (and correct) proof?
    \item[] Answer: \answerNA{} 
    \item[] Justification: The paper does not include theoretical results.  

    \item {\bf Experimental result reproducibility}
    \item[] Question: Does the paper fully disclose all the information needed to reproduce the main experimental results of the paper to the extent that it affects the main claims and/or conclusions of the paper (regardless of whether the code and data are provided or not)?
    \item[] Answer: \answerYes{} 
    \item[] Justification: We have detailed data construction process, model settings, and training process in \Cref{app:implementation} to ensure the reproducibility of our results.

\item {\bf Open access to data and code}
    \item[] Question: Does the paper provide open access to the data and code, with sufficient instructions to faithfully reproduce the main experimental results, as described in supplemental material?
    \item[] Answer: \answerYes{} 
    \item[] Justification: We have uploaded the code to an anonymous link in the paper.

\item {\bf Experimental setting/details}
    \item[] Question: Does the paper specify all the training and test details (e.g., data splits, hyperparameters, how they were chosen, type of optimizer, etc.) necessary to understand the results?
    \item[] Answer: \answerYes{} 
    \item[] Justification: We have detailed experiment settings in \Cref{app:implementation}.

\item {\bf Experiment statistical significance}
    \item[] Question: Does the paper report error bars suitably and correctly defined or other appropriate information about the statistical significance of the experiments?
    \item[] Answer: \answerNA{} 
    \item[] Justification: The experiments are conducted with a fixed random seed and no error bars are reported.

\item {\bf Experiments compute resources}
    \item[] Question: For each experiment, does the paper provide sufficient information on the computer resources (type of compute workers, memory, time of execution) needed to reproduce the experiments?
    \item[] Answer: \answerYes{} 
    \item[] Justification: The resources used for the experiments are detailed in \Cref{app:training_settings}.
    
\item {\bf Code of ethics}
    \item[] Question: Does the research conducted in the paper conform, in every respect, with the NeurIPS Code of Ethics \url{https://neurips.cc/public/EthicsGuidelines}?
    \item[] Answer: \answerYes{} 
    \item[] Justification: The research conducted in the paper complies with the NeurIPS Code of Ethics.

\item {\bf Broader impacts}
    \item[] Question: Does the paper discuss both potential positive societal impacts and negative societal impacts of the work performed?
    \item[] Answer: \answerNA{} 
    \item[] Justification: The proposed method focuses on the technical aspects of the problem and do not include societal impacts.
    
\item {\bf Safeguards}
    \item[] Question: Does the paper describe safeguards that have been put in place for responsible release of data or models that have a high risk for misuse (e.g., pretrained language models, image generators, or scraped datasets)?
    \item[] Answer: \answerNA{} 
    \item[] Justification: The paper utilizes existing datasets and pretrained models that are already released which have safeguards in place. 

\item {\bf Licenses for existing assets}
    \item[] Question: Are the creators or original owners of assets (e.g., code, data, models), used in the paper, properly credited and are the license and terms of use explicitly mentioned and properly respected?
    \item[] Answer: \answerYes{} 
    \item[] Justification: We have properly cited all code, data, and models used in our research and complied with the licensing agreements and terms of use set by the original authors.

\item {\bf New assets}
    \item[] Question: Are new assets introduced in the paper well documented and is the documentation provided alongside the assets?
    \item[] Answer: \answerNA{} 
    \item[] Justification: There are no new assets introduced in the paper.

\item {\bf Crowdsourcing and research with human subjects}
    \item[] Question: For crowdsourcing experiments and research with human subjects, does the paper include the full text of instructions given to participants and screenshots, if applicable, as well as details about compensation (if any)? 
    \item[] Answer: \answerNA{} 
    \item[] Justification: The paper does not involve crowdsourcing nor research with human subjects.

\item {\bf Institutional review board (IRB) approvals or equivalent for research with human subjects}
    \item[] Question: Does the paper describe potential risks incurred by study participants, whether such risks were disclosed to the subjects, and whether Institutional Review Board (IRB) approvals (or an equivalent approval/review based on the requirements of your country or institution) were obtained?
    \item[] Answer: \answerNA{} 
    \item[] Justification: The paper does not involve crowdsourcing nor research with human subjects.

\item {\bf Declaration of LLM usage}
    \item[] Question: Does the paper describe the usage of LLMs if it is an important, original, or non-standard component of the core methods in this research? Note that if the LLM is used only for writing, editing, or formatting purposes and does not impact the core methodology, scientific rigorousness, or originality of the research, declaration is not required.
    \item[] Answer: \answerYes{} 
    \item[] Justification: The usage of the LLM is described and discussed in \Cref{app:implementation,app:kg_index}.

\end{enumerate}

\end{document}